\documentclass[fleqn,usenatbib]{mnras}

\usepackage{newtxtext,newtxmath}

\usepackage[T1]{fontenc}

\DeclareRobustCommand{\VAN}[3]{#2}
\let\VANthebibliography\thebibliography
\def\thebibliography{\DeclareRobustCommand{\VAN}[3]{##3}\VANthebibliography}

\usepackage{graphicx}	
\usepackage{amsmath}	
\usepackage{graphicx} 
\usepackage{float}
\usepackage{array}
\usepackage{multirow}
\usepackage{booktabs}
\usepackage{csquotes}
\usepackage{subcaption}
\usepackage{adjustbox}
\usepackage{lipsum}
\usepackage{siunitx}
\usepackage{gensymb}
\usepackage{calc, graphicx}  
\usepackage{pdfpages}
\usepackage{verbatim}
\usepackage{booktabs}
\usepackage{multirow}
\usepackage{geometry}
\usepackage{longtable}
\usepackage{placeins}
\usepackage{booktabs}    
\usepackage{geometry}    
\usepackage{ragged2e}    
\usepackage{pdflscape} 
\usepackage{longtable} 
\usepackage{geometry} 
\usepackage{adjustbox} 
\usepackage{makecell}

\renewcommand{\arraystretch}{1} 

\title[Galactic X-ray Transients in eRASS1]{Galactic X-ray Transients in the First eROSITA All Sky Survey}

\author[Maan et al.]{
Vikash Maan,$^{1}$\thanks{E-mail: vikashmaan98@gmail.com}
Aman Katira,$^{1}$
Kunal. P. Mooley$^{1,2}$
\\
$^{1}$Indian Institute Of Technology Kanpur, Kanpur, Uttar Pradesh 208016, India\\
$^{2}$Caltech, 1200 E. California Blvd.  MC 249-17, Pasadena, CA 91125, USA}

\date{Accepted XXX. Received YYY; in original form ZZZ}

\pubyear{\the\year{}}

\begin{document}
\label{firstpage}

\pagerange{\pageref{firstpage}--\pageref{lastpage}}
\maketitle

\begin{abstract}
Although a multitude of studies have focused on targeted observations of Galactic X-ray transients, blind surveys and population studies have been limited.
We have used the \textit{ROSAT}, eROSITA and Gaia source catalogs to find Galactic X-ray transients having timescales $<$30 years. We report the properties of 738 transients found in our search, majority of which are active stars or interacting binaries. 
We have also found $\sim$40 compact object systems among which are at least 8 newly-identified white-dwarf systems, 3 known X-ray binaries, and one known pulsar.
We use eROSITA (soft X-ray) spectra of the Galactic transients to show that two distinct types of flaring systems are prevalent: one having peak around 1 keV, well fit by thermal models, and another having peak below 0.2 keV and requiring a power-law component.
Our study also reveals that single star or interacting binary systems (X-ray transients) involving giant stars exhibit significantly higher X-ray luminosities than systems involving only main-sequence stars or young stellar objects.
Finally, we discuss the properties of the transients in the context of their putative emission mechanisms, the fraction of transients with respect to the total population, and the rates of Galactic transients expected in blind searches of the X-ray sky.  
\end{abstract}

\begin{keywords}
X-rays: binaries, stars: cataclysmic variables, stars: supergiants
\end{keywords}



\section{Introduction}
Galactic X-ray transients\footnote{In this work we refer to variables and transients collectively as "transients".} span a range of classes, including chromospherically active main-sequence stars (single and binary; flaring events in F-, G-, K-type stars and M dwarfs, RS CVn, BY Dra-type and contact W UMa-type systems), flaring Young Stellar Objects (YSOs) owing to their enhanced magnetic activity, outbursts in accreting compact object systems (e.g. Thermonuclear bursts and hydrogen ionization instability related outbursts in X-ray binaries, classical and dwarf novae in cataclysmic variables) as well as wind-driven activity related to massive and evolved stars. 
Each of these classes of X-ray emitting stars/stellar systems have been studied in some detail through targeted observations \citep[e.g.][]{1980_carroll_W_UMa,1980_white_algols,RS_CVn_walter_bowyer_1981,1983_white_marshall_alogols,Cruddace_1984_W_UMa,Drake_1989_RS_CVn,1993_dampsey_RS_CVn,1994_fox_RS_CVn,Warner_1995,alpha_square_dynamo_effect_Giampapa,1996_KP_SINGH_Alogols,Fender_2001_XRB_Review,Gudel_and_manual,Remillard_et.al_2006,Gudel_et.al,Drake.et.al_2019}.

A diverse range of emission mechanisms are known to be associated with all of these Galactic X-ray transients. Ubiquitous flaring phenomena, attributed to magnetic reconnection, magnetically-accelerated plasma, and/or increased collisional excitation/de-excitation (emitting via thermal/non thermal bremsstrahlung or free bound transitions), are thought to be the cause of transient X-ray emission in YSOs and low-mass main-sequence stars \citep[e.g][]{Ozawa_et_al,YSO_flare_NuSTAR,yso_flare,Flaring_yso_xmm_obs,Zhuleku_et.al,Drake.et.al_2019,Gudel_et.al}. 
Interacting binary systems (non-accreting; primarily Algol- and RS CVn-type systems), which exhibit frequent X-ray variability, have undergone speeding up of rotation due to gravitational interaction thereby resulting in enhanced chromospheric activity \citep[e.g.][]{universe_in_x_rays} of one or both components of the binary.
High-mass single stars may also exhibit variability that may be attributed to variations in stellar wind collisions with the circumstellar media that lead to the production of varying degrees of shocks \citep[e.g.][]{Lucy_and_white}.

Accreting compact objects are another important class of Galactic X-ray transients. 
Cataclysmic variables (CVs), further classified into dwarf novae \citep[e.g.][]{Lasota_2001,dwarf_novae_1,dwarf_novae_2}, classical novae \citep[e.g.][]{universe_in_x_rays,CLASSICAL_NOVAE_1,CLASSICAL_NOVAE_2}, symbiotic systems \citep[e.g.][]{Drake.et.al_2019,Warner_1995}, and other classes, can produce transient X-ray emission owing to disk instability-induced outbursts, runaway thermonuclear fusion reactions or simply accretion state changes \citep[e.g.][]{symbiotic_systems_review,symbiotic_systems_x_ray_emission_mechanism,Mukai_review_of_CV_and_symbiotic_systems}. Transient X-ray emission from these systems is believed to be primarily thermal (blackbody or thermal bremsstrahlung).
X-ray binary (XRB) systems in outburst are also sources of copious X-ray flares. 
In XRBs the flaring results from thermonuclear bursts (for neutron star systems), accretion state changes or jets \citep[e.g][]{Fender_2001_XRB_Review,Fender_2006_XRB_Jets,XRB_review,Z_atoll_source_review,Remillard_et.al_2006,Muno_et_al,2007_done_XRB_Outburst}. 
Transient emission in soft X-rays is due to optically thick emission from the accretion disk surrounding a compact object, while hard X-rays originate from the inverse Compton process in the corona \citep[e.g.][]{Fender_2001_XRB_Review}.


Several blind searches for Galactic X-ray transients, employing the XMM-Newton \citep[e.g][]{page_xmm_cataloge,webb_xmm_cataloge,2009_wetson_XMM_serendipitous_survey}, Chandra \citep[e.g][]{evans_chandra_cataloge,evans_2_chandra_cataloge} and \textit{ROSAT} source catalogs \citep[e.g][]{rosat_all_sky_survey,Boller2016}, have been executed previously. \citealt{XMM_vs_ROSAT_2022_SURVEY} cross-matched the XMM-Newton slew survey and \textit{ROSAT} catalogs to find: a) that $\sim$2.5\% of all the Galactic and extra-Galactic X-ray sources in the sky are variable, b) among single stars (not interacting binaries), cooler stars exhibit higher variability then  hotter stars, and c) high mass X-ray binaries are significantly more variable than low mass X-ray binaries.
\citealt{YUAN_2006} undertook a two epoch variability search between \textit{ROSAT} and XMM-Newton Serendipitous Survey catalogs, which resulted in 10 sources, including a CV, a microquasar, four flare stars, an
AGN and rest unidentified sources, having  XMM to \textit{ROSAT} flux ratio greater than 10 between the two surveys. \citealt{X_ray_sources_classified_from_XMM_data} classified unknown X-ray transients in XMM-Newton Serendipitous Source Catalog as stars, AGN and compact objects, finding that among different categories of sources, X-ray binaries exhibit large variability while flaring stars show the least variability. \citealt{Kaya_Mori_2021} used Chandra data of the Galactic center to find that 25\% of the X-ray binary population shows transient behavior over a span of 16 years.

Some previous works have attempted to carve out a luminosity-timescale phase space for Galactic X-ray transients. 
Flaring stars (single stars and non-accreting interacting binaries) have peak X-ray luminosities of $L_X \lesssim10^{33}$ erg s$^{-1}$, and variability timescales between $\sim$minutes and hours ; $L_X \sim 10^{29-37}$ erg s$^{-1}$ and variability timescales between days and weeks are associated with cataclysmic variables; $L_X\sim 10^{36-39}$ erg s$^{-1}$ and variability timescales between sub-second and days for X-ray binaries ($L_X\sim 10^{30-34}$ erg s$^{-1}$ in quiescence; \citealt{Remillard_et.al_2006})  \citep[e.g][]{yso_flare,Soderberg_2009,Mukai_review_of_CV_and_symbiotic_systems,CV_luminosity,lmxb_luminosity,symbiotic_systems_x_ray_emission_mechanism}.

Although targeted surveys have been able to probe low-luminosity sources, there is a modicum of sensitive blind searches to probe Galactic transients below $L_{X}\simeq10^{30}$ erg s$^{-1}$.
The X-ray sky survey with the extended ROentgen Survey with an Imaging Telescope Array aboard Spektrum-Roentgen-Gamma (SRG) observatory \citep[e.g][]{2021_predehl_erosita_telescope} gives us an unprecedented opportunity to execute exactly such a study. 
eROSITA's excellent sensitivity (flux limit $\sim 5\times10^{-14}$ erg s$^{-1}$ cm$^{-2}$  , in the 0.5--2 keV energy band for eROSITA All Sky Survey 1 (eRASS1); \citealt{Merloni2024}) enables the study of low-luminosity transients. The availability of X-ray spectral products facilitates spectral analysis; and blind search across a wide field (2$\pi$ sr of the sky) allows the investigation of transient rates and simultaneous study of populations of transient classes. In this paper we compare eRASS1 \citep[e.g][]{Merloni2024}, the second \textit{ROSAT} All-Sky Survey Point Source (2RXS) catalog \citep[e.g][]{Boller2016} and the Gaia DR2 \citep[e.g][]{GaiaCollaboration2018} source catalogs to study Galactic X-ray transients on timescales between a few hours and 30 years.

This paper is organized as follows. In Section 2 we describe the X-ray data, optical data, and the cross-matching of the X-ray and optical source catalogs. In Section 3 we describe the identification and classification of X-ray transient sources. In Section 4, we outline the important X-ray luminosity and spectral properties of Galactic transients, as well as the characteristics of nine new potential CVs, three known XRBs, one known pulsar and Wolf Rayet detected in our analysis. We conclude with a summary and discussion in Section 5.

\section{X-ray Data, Optical Data and Data Processing}

\subsection{The eRASS1, 2RXS and Gaia DR2 catalogs}

We used the eRASS1 catalog which contains the X-ray source positions, as well as flux and source extent information (energy range 0.2--2.3\,keV) for 0.93 million sources observed during eROSITA's first all sky survey executed between December 2019 and June 2020. It spans 2$\pi$ steradians( \citealt{Merloni2024}).
In this catalog, the reported absorbed flux in the 0.2--2.3\,keV band folds in a powerlaw model having photon index $\Gamma=2$ \citep[e.g][]{Merloni2024}.

The 2RXS catalog (\citealt{Boller2016}) contains information about 0.13 Million X-ray sources detected during the all sky survey undertaken between June 1990 and August 1991 in the energy range 0.1-2.4 keV. 
The 2RXS catalog contains count rates and absorption-corrected flux where a different hydrogen column density and $\Gamma$ have been used for each source depending on its position and nature \citep[e.g][]{Boller2016}.
We therefore converted the \textit{ROSAT} count rates into absorbed flux in the  eROSITA energy band with photon index 2 using WebPIMMS\footnote{\url{https://heasarc.gsfc.nasa.gov/cgi-bin/Tools/w3pimms/w3pimms.pl}}. 
Accordingly, we find the \textit{ROSAT} count rate-to-flux conversion factor (energy conversion factor, ECF) to be $9.04\times10^{10}$ cm$^2$ erg$^{-1}$. \footnote{We followed this process in order to find the 2RXS flux in exactly the same way it has been done for eRASS1 sources \citep[e.g][]{Merloni2024} . In this work, we note that we have found 2 main types of transient source X-ray (eRASS1) spectra: (1) spectra peaking at around 1keV ($\sim$700 transients)  (2) steep spectra (31 transients) (please see text for details). If we take these spectral shapes into account when calculating the fluxes (F$_{\rm spec}$), then we find that the F$_{\rm spec}$ of sources having peaked spectra agrees with the F$_{\rm eRASS1}$ (calculated using $\Gamma$=2) within their respective $1\sigma$ uncertainties, while the F$_{\rm spec}$ of sources having steep spectra ($\Gamma$ found to vary between 3 to 7) are, on an average, a factor of two larger than F$_{\rm eRASS1}$.}

\begin{figure}
    \centering
    \includegraphics[scale=0.45]{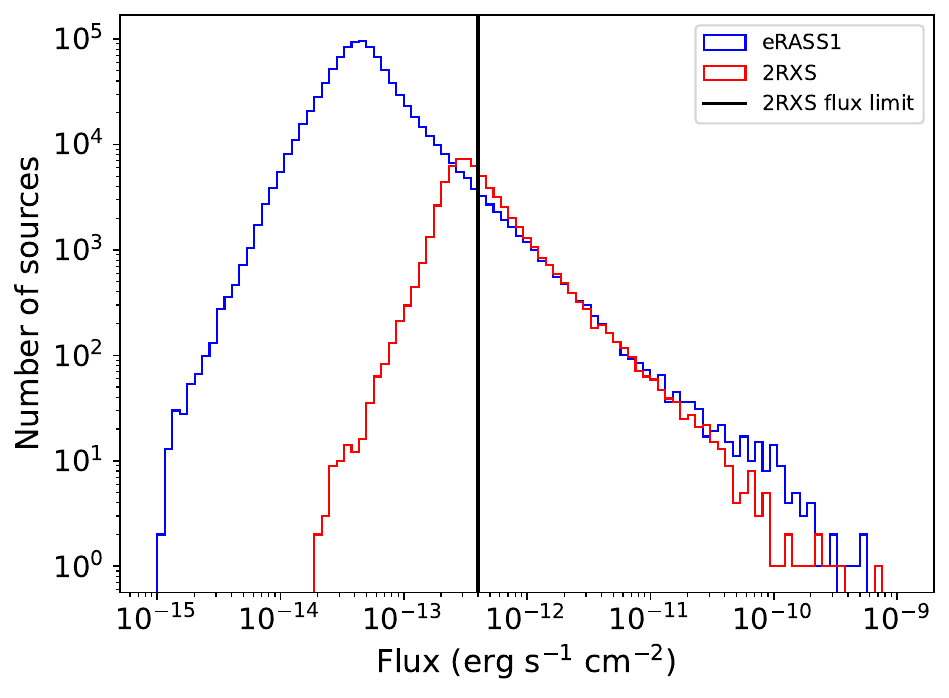}
    \caption{Histograms of the flux values (absorbed, assuming photon index 2) of 2RXS and eRASS1 catalogs. The approximate 2RXS flux threshold limit is marked with the vertical line at $4\times10^{-13}$ erg  s$^{-1}$ cm$^{-2}$}
    \label{fig:flux-hist}
\end{figure}

In Figure~\ref{fig:flux-hist}, we show histograms of the 2RXS and eRASS1 fluxes, both using photon index 2, absorbed, and in the 0.2--2.3\,keV energy band. 
We find approximate flux limits for the eRASS1 and 2RXS catalogs, i.e. close to the peaks of the flux histograms, as $F_{\rm lim,eRASS1} = 5\times10^{-14}$ erg  s$^{-1}$cm$^{-2}$ and $F_{\rm lim,2RXS} = 4\times10^{-13}$erg  s$^{-1}$cm$^{-2}$ respectively.\footnote{cgs to SI units conversion: 1 erg s$^{-1}$ cm$^{-2}$ = 10$^{-3}$ W m$^{-2}$}

We also used the Gaia DR2 catalog to get the precise parallaxes, positions, as well as the photometric magnitudes G, BP and RP, of Galactic optical sources for 1.7 billion sources. We also used the effective temperature, stellar radius, and reddening information from Gaia DR2 wherever available (this information exists for $\sim$120 million sources).

\subsection{eROSITA X-ray Spectral Products}\footnote{In this work we have used eROSITA spectral products to analyze the properties of the transients, not to calculate fluxes and luminosities. These latter quantities are based on the published eRASS1 catalog and Gaia distances.} 
eROSITA X-ray spectral data (response files -- ARF and RMF -- as well as source and background spectra) were obtained from their science data products website\footnote{\url{https://eROSITA.mpe.mpg.de/dr1/erodat/catalogue/search_by_id}}. 
We used the X-ray spectral fitting tool XSPEC (version 12.13.1;  \citealt{Arnaud1996}) available in HEASoft (version 6.32.1) to extract the background-corrected X-ray spectra.
For each source of interest, we used the part of the spectral files corresponding to the combined data from all 7 eROSITA modules in the energy range 0.2--10 keV. 
To increase the signal-to-noise ratio we used \textit{grppha} to group the spectral data points to 5 counts per bin.
The resulting spectra were plotted using PyXspec and will be discussed in the following sections. 

\subsection{Gaia multi-epoch photometric data} 
We used Gaia DR3 \citep[e.g][]{GaiaCollaboration2023} to obtain photometric time series data in BP, RP and G bands. 
We used these data to find the periods associated with individual eRASS1 sources. 
We used \textit{getspy} adopted by \textit{astropy} to get power density spectrum (plot of power vs frequency) using Lomb-Scargle periodogram (LS periodogram, e.g.  \citet{Lomb1976,Scargle1982}). LS periodogram was used to find the period associated with the unevenly sampled time series data.

\subsection{eRASS1 and Gaia DR2 cross-match}
We selected 20655 sources from eRASS1 which satisfy the flux condition $F_{\rm eRASS1} > F_{\rm lim,2RXS}$, where $F_{\rm eRASS1}$ represents the eRASS1 flux.
Since in this work we are focusing on Galactic transients, we cross-matched the 20655 eRASS1 sources with the Gaia DR2 catalog. 
We further shortlisted the matched sources based on the following criteria: (i) the nearest cross-matched source is considered as the optical counterpart, (ii) the Gaia source parallax is positive and finite, and (iii) the parallax signal-to-noise ratio is $>$10.

The procedure resulted in 2916 common sources between the two catalogs. 
For this exercise we used a 2" matching radius, justified as follows. 
In Figure~\ref{fig:match-radius} (dark blue dashed line) we show the fraction of matched sources between the two source lists/catalogs and matching radius.
Below about 2'' we find a constant increase in the number of matched sources (as a function of matching radius), while above this radius we find a change in the rate of increase of matched sources.
Given the positional uncertainties of the eRASS1 sources (2'') and Gaia sources ($\ll$1''; \citealt{GaiaCollaboration2023}), we chose a matching radius \footnote {Some previous studies like \cite{2024_Rodriguez_X_ray_main_sequence} 
 have also used 2" as the matching radii between 4XMM-DR13 and Gaia DR3.} of 2''. 
To estimate the false-positives among the matched sources we changed all eRASS1 sources co-ordinates by 1' {\footnote {All eRASS1 sources have been systematically shifted in the same direction by 1'} and repeated the cross-matching exercise.
 The resulting curve is shown as a light blue dashed line in Figure~\ref{fig:match-radius}. 
We interpret this as the false-positive detection rate for the eRASS1-Gaia DR2 match of <15\% at 2" matching radius (green solid line in Figure~\ref{fig:match-radius}).

\subsection{eRASS1 and \textit{ROSAT} (2RXS) cross-match}

We cross-matched 20655 sources with 2RXS\footnote{Available on Vizier with CDS code J/A+A/588/A103/cat2rxs} 
with different matching radii using the CDS tool X-match\footnote{\url{http://cdsxmatch.u-strasbg.fr/}}. 
The result of this exercise is shown with the red dotted curve in Figure~\ref{fig:match-radius}, as the fraction of eRASS1 sources, that have a match in 2RXS, versus the matching radius. 
Following the same procedure as with the eRASS1 and Gaia DR2 cross-match, we changed the eRASS1 coordinates by 1' and then repeated the cross-match with 2RXS.
This result is shown with the magenta dotted line in Figure~\ref{fig:match-radius}.
The \textit{ROSAT} source population density of 2.4 sources deg$^{-2}$ and a matching radius of 40'' implies a false-positive rate of 0.1\% in the list of matched sources.
We also note that the median positional uncertainty in the eRASS1 source coordinates is 2'', while that for the \textit{ROSAT} source coordinates\footnote{We could not locate the median positional uncertainties for sources in the 2RXS catalog, hence we use the 1RXS catalog \citep[e.g][]{1999_rosat_bright_source_cataloge_1rxs} to estimate the median positional uncertainty} is 11''.
Some previous works \citep[e.g.][]{XMM_vs_ROSAT_2022_SURVEY,YUAN_2006} have used 1'--2' as the radius for cross-matching X-ray sources with the \textit{ROSAT} catalog.
Taken together (along with the precision in the source coordinates of the two catalogs; see Sec. 2.1), we conclude 40'' to be a reasonable matching radius between eRASS1 and 2RXS since there is no substantial incremental gain, in terms of the number of matched sources, beyond this radius and that the number of chance coincidences is $\ll$10\%.
Using the 40'' cross-matching radius we find 10671 common sources between the eRASS1 and 2RXS catalogs and 9984 sources unique to eRASS1. 

For the common sources we plot the ratio of the 2RXS and eRASS1 fluxes in Figure~\ref{fig:Gaussian fitted flux ratios}. The histogram nicely follows a Gaussian distribution having a fitted mean of $1.15\pm 0.01$ and standard deviation $\sigma=1.86\pm0.01$.
 We adopted the following procedure to identify the error bars on the bins shown in the Figure~\ref{fig:Gaussian fitted flux ratios}. For every source, we used 2RXS and eRASS1 flux$\pm$uncertainty to generate 100 sets of random numbers for the flux ratio \footnote{We used numpy.random.uniform in python to generate the random numbers.}. For each set, a histogram was constructed using identical binning, resulting in an ensemble of 100 histograms. The spread of the counts in each bin across this ensemble was then quantified, and the standard deviation was adopted as the statistical uncertainty for that bin in the Figure ~\ref{fig:Gaussian fitted flux ratios}.
 
We corrected 2RXS fluxes for the 15\% flux offset during the variable source search described in the next section\footnote{
{Flux offset corrections have been widely used in previous variability/transient searches, \citep[e.g][]{Mooley_flux_offset_correction_2016,2011_Thyagarajan_flux_offset_correction}.
In order to investigate the source of flux offset in our work, we selected a series of samples brighter than $F_{\rm lim,2RXS}$. For these brighter samples we find that the flux ratio is centered around unity. Thus we find that for sources significantly brighter than $F_{\rm lim,2RXS}$, the flux offset goes away. This likely suggests Eddington bias as the source of the offset. If we correct for flux-dependent offsets, then the number of transients identified in our work reduces by $\sim$10. Since this is a relatively small change in the total number of transients found in this work, we have retained a single correction factor for all sources, for simplicity.}}.

 \begin{figure}
    \centering
    \includegraphics[scale=0.45]{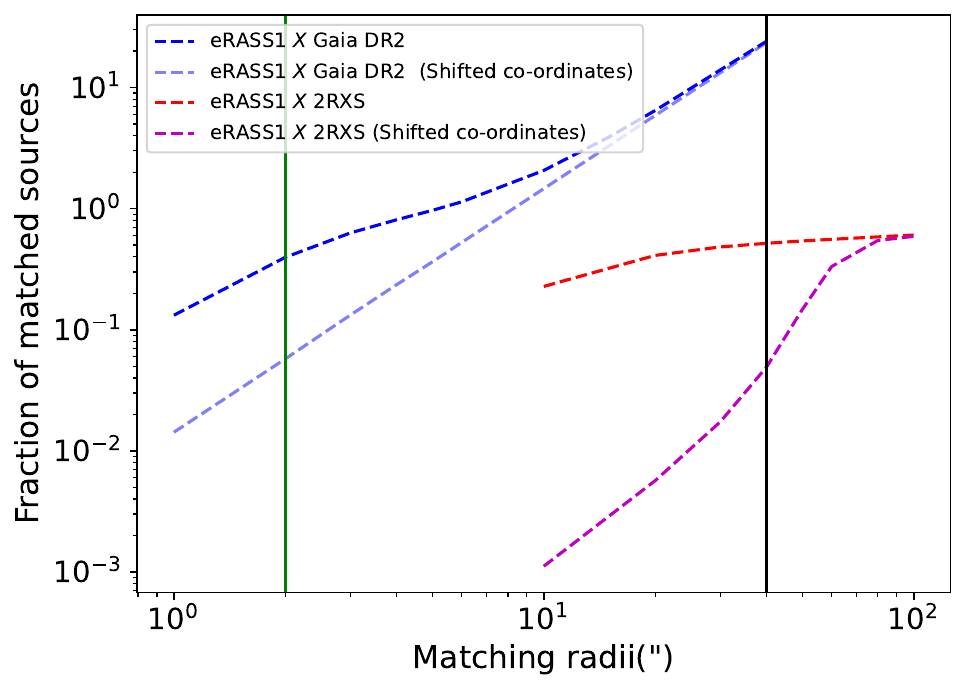}
    \caption{Plot showing the matching radius versus the fraction of matched sources (total cross-matched sources/20655) for the cross-match of eRASS1 with Gaia DR2 and 2RXS catalogs. We chose 40" and 2" (black and green vertical line respectively) as the matching radius for eRASS1/2RXS cross-match and eRASS1/Gaia DR2 cross-match respectively. Magenta and light blue data points correspond to the above mentioned cross-matches but with eRASS1 sources shifted by 1'.}
    \label{fig:match-radius}
\end{figure}
 


\begin{figure}
    \centering
    \includegraphics[scale=0.25]{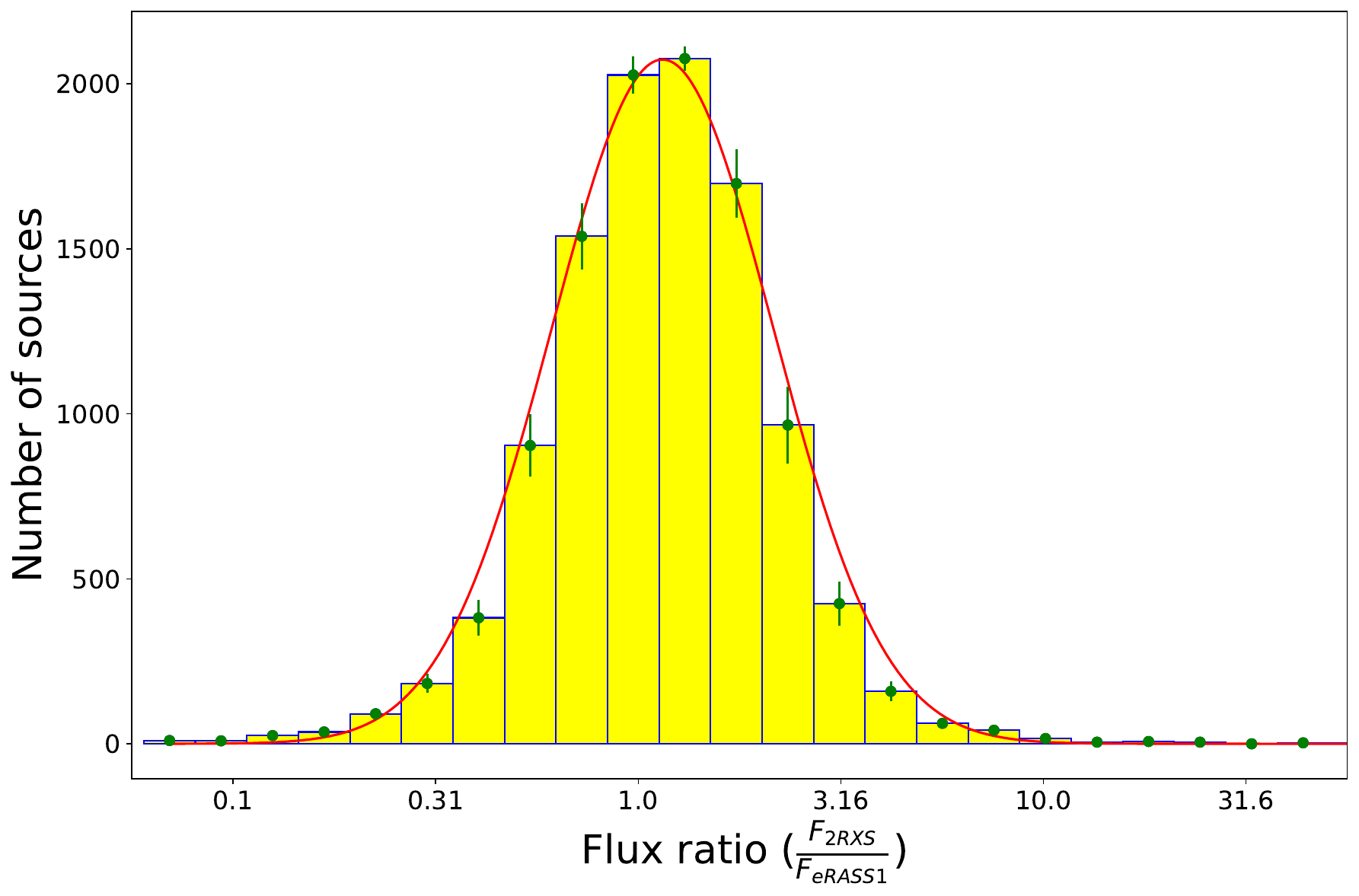}
    \caption{Histogram of the 2RXS-to-eRASS1 flux ratio for the 10671 sources resulting from the cross-match of eRASS1/2RXS using a 40'' matching radius. The error bars on each bin are described in the text. The red curve represents a Gaussian fit to the histogram and indicates a mean value (for the flux ratio) of $1.15\pm 0.01$ and a corresponding standard deviation of $1.86\pm0.01$. }
    \label{fig:Gaussian fitted flux ratios}
\end{figure}  

 \begin{figure}
    \centering
    \includegraphics[scale=0.9]{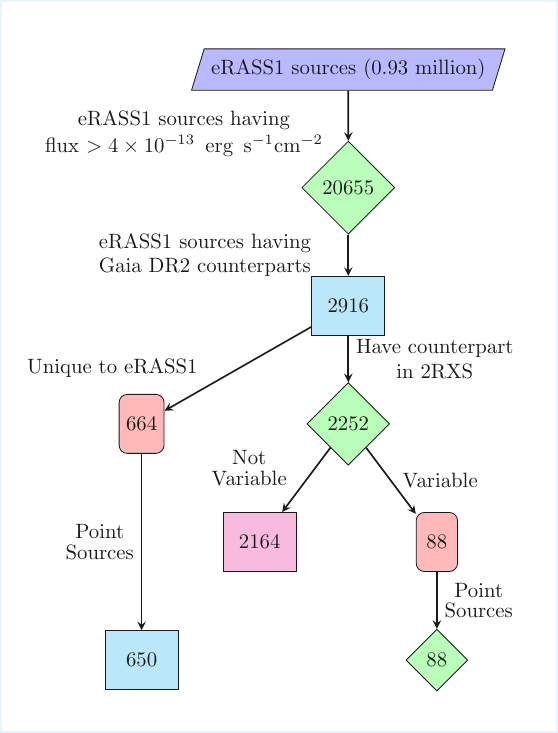}
    \caption{Flowchart showing the procedure for selecting Galactic X-ray transients in the eRASS1 catalog.}
    \label{fig:eROSITA_flowchart}
\end{figure}

   
\vspace{0.3cm}
\vspace{0.3cm}
\vspace{0.3cm}
\vspace{0.3cm}

\section{ Galactic Transient search and  Source Classification}\label{sec:transients search}

We cross-matched the 20655 eRASS1 sources (discussed in Sec. 2.4) with Gaia DR2 to get 2916 matched sources, which were further cross-matched with the 2RXS catalog.
This resulted in 2252 sources matched between the three catalogs and 664 sources that are absent in 2RXS.
We then proceeded with transient search as described below. The full procedure is also shown as a flowchart in Figure~\ref{fig:eROSITA_flowchart}.

In this work we refer to variables (i.e. sources detected in both, the 2RXS and eRASS1, catalogs but with significantly different fluxes, taking into account the flux offset as discussed in the previous section) and true transients (detected in eRASS1 but not in 2RXS) together as "transient sources". 

The first part of our transient search entails the selection of sources (variables) that are detected in both eRASS1 and 2RXS catalogs, but are $2\sigma$ on either side of the mean value of the ratio of fluxes plotted in Figure~\ref{fig:Gaussian fitted flux ratios}.
This gives us 88 transients. 
The variability significance for these sources (transients), quantified through the statistic  $V_S = |F_{\rm 2RXS} - F_{\rm eRASS1}|/\sqrt{\sigma_{\rm 2RXS}^2 + \sigma_{\rm eRASS1}^2} = \Delta F/\sigma$, was found to be $\gtrsim$3, indicating that all the sources are significant variables.

The second part of our transient search entails the selection of point-like sources that are unique to eRASS1 (i.e. not detected in the 2RXS catalog). This gives us 650 additional transients. 

In summary, among the 2916 Galactic X-ray sources, we find 738 transient sources. The important parameters associated with these sources are given in Table~\ref{Transients table}. 
Their sky locations are given in Figure~\ref{fig:lat-lon} and the luminosity-distance 
plot for the sources are shown in Figure~\ref{fig:LVD}.  

\begin{figure*}
\includegraphics[width=0.9\textwidth]{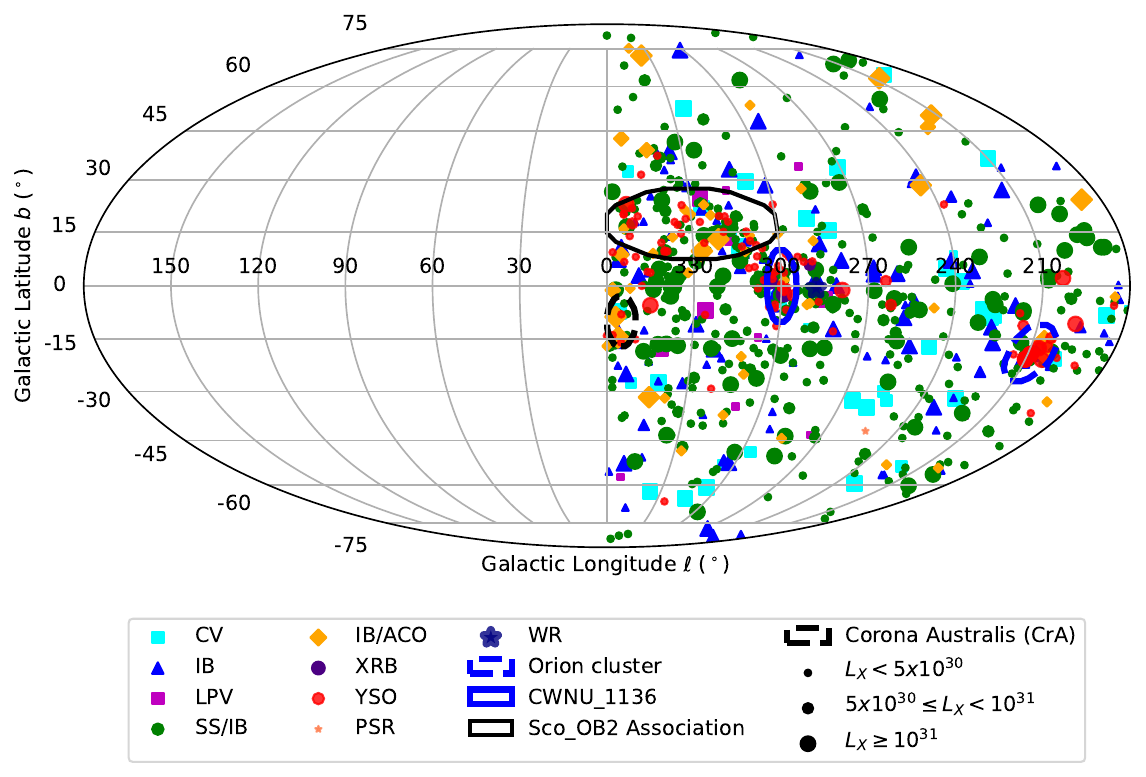}
\caption{Locations of Galactic transients in Galactic co-ordinates, highlighting the open clusters and molecular clouds to which YSOs belong. Various classes of Galactic transients mentioned in section 3.1 are color coded and the varying marker correspond to the different X-ray luminosities (in erg s$^{-1}$) of the transients, as shown in the legend. }
\label{fig:lat-lon}   
\end{figure*}

\begin{figure}
\includegraphics[width=0.5\textwidth]{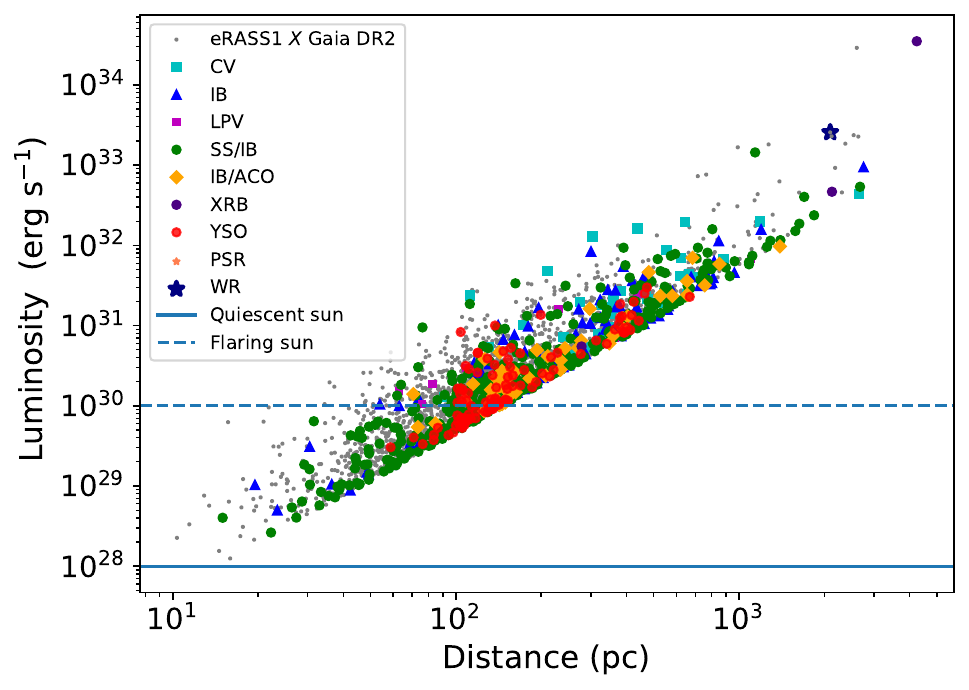}
\caption{Luminosity (in the 0.2--2.3 keV energy band) vs distance plot for Galactic  transients in eRASS1, plotted on the log-log scale. Grey data points correspond to eRASS1/Gaia DR2 cross-match .Various classes of Galactic transients mentioned in section 3.1 are color coded and the $L_X$ of quiescence Sun is from \citealt{Sun_X_ray_luminosity}.}
\label{fig:LVD}
\end{figure}


\subsection{Source Classification}
We have classified the 738 transient into the following categories, as given in Table~\ref{Transients table}.
\begin{enumerate}
\item YSO: Young stellar object (105 Transients)
    \item SS/IB : Either chromospherically active single star or interacting binary (418 Transients)
    \item IB: Known interacting binary not having any compact object in the system (110 Transients)
    \item IB/ACO: Interacting binary or accreting compact object (50 Transients)
    \item CV : Cataclysmic variable (38 Transients)
    \item XRB : X-ray binary (3 Transients)
    \item LPV :  Long period variable (12 Transients)
    \item PSR : Pulsar (1 Transient)
    \item WR : Wolf Rayet (1 Transient) 
\end{enumerate}

We used several criteria to classify the X-ray transients, including X-ray luminosities ($L_X$), X-ray spectra, positions in the Hertzsprung-Russel (HR) diagram, positions relative to the "X-ray main sequence" \citep[e.g][]{2024_Rodriguez_X_ray_main_sequence} and classification in databases like {\it Simbad} \citep[e.g][]{2000_wegner_Simbad}.
Our classification procedure is detailed below.

\begin{enumerate}
\item We queried the {\it Simbad} database for classifications of the X-ray transient sources using {\it astroquery}. 
{\it Simbad} classifications such as RS CVn, Spectroscopic Binary, Eclipsing Binary, Double or Multiple Star were labeled as "IB" in our analysis.
Classifications like Young Stellar Object (108 sources) and T Tauri Star (16 sources) were labeled as "YSO". Transients classified as Cataclysmic Binary/Hot Subdwarf Candidate (29 sources) and X-ray Binary (3 sources) were labeled as "CV" and "XRB" respectively. Additionally, transients classified as Long Period Variable (18 sources), Pulsar (1 source) and Wolf Rayet (1 source) were labeled as "LPV", "PSR" and  "WR" respectively. Remaining classified transients were labeled as "SS/IB", we did not label any transient as "SS" (Single star) because its binarity information may be incomplete within {\it Simbad}.
About $\sim$100 transients did not have any {\it Simbad} classification. 

\begin{figure*}
    \centering
        \centering
        \includegraphics[scale=0.8]{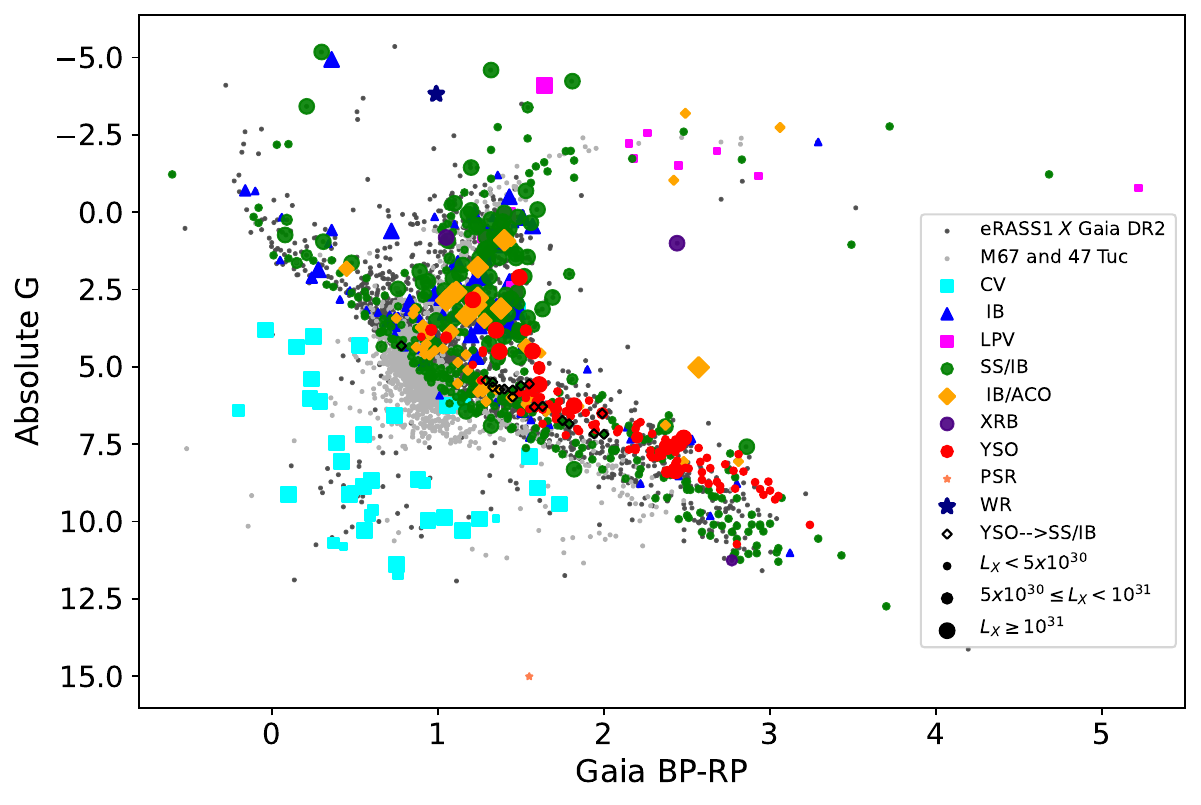} 
        \centering
        \includegraphics[scale=0.8]{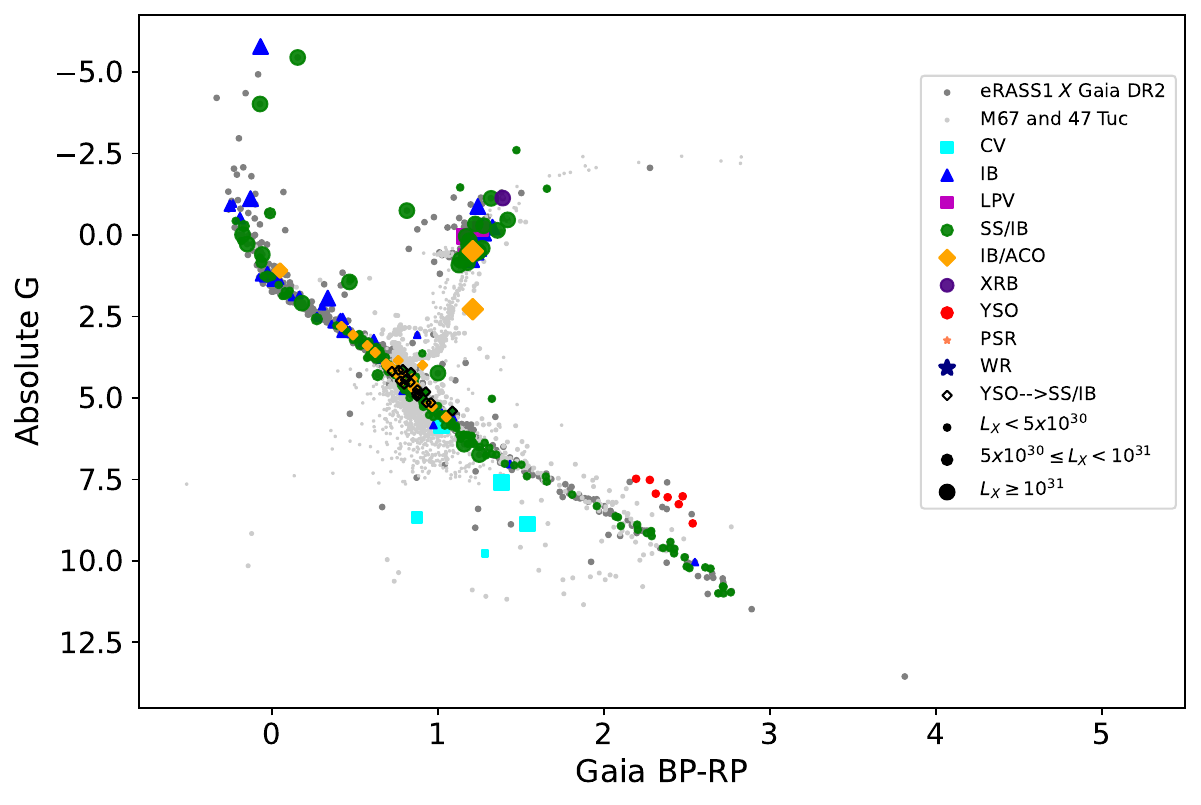} 
        \caption{The HR diagrams for the Gaia counterparts of the Galactic X-ray transients in eRASS1. Light grey points in the background correspond to the M67 and 47Tuc stellar clusters respectively. Dark grey data points correspond to eRASS1 sources having a Gaia DR2 counterpart. Various classes of Galactic transients mentioned in section 3.1 are color coded.  \textit{Upper Panel}: Not corrected for reddening (Total 722 transients). \textit{Lower Panel}: Corrected for reddening. Only those transients having extinction information have been plotted in the \textit{lower panel} (303 transients). Varying marker correspond to the different X-ray luminosities (in erg s$^{-1}$) of the transients, as shown in the legend. Transients shown with empty black diamond are classified as Young Stellar Object or T Tauri Star in \textit{Simbad}, but they all lie along the main sequence branch in the lower panel. Consequently, we reclassified them into "SS/IB" category.} 
    \label{fig:HRD}
\end{figure*}

\item Transients whose Gaia counterparts lie below the main sequence on the HR diagram (Figure~\ref{fig:HRD}) were labeled as "CV" in our analysis, as this region of the HR diagram is occupied by systems containing white dwarfs. 
To further verify the "CV" classification, we investigated the locations of the transients on the "X-ray main sequence" diagram shown in Figure~\ref{fig:x-ray-main-sequence}  \citep[e.g][]{2024_Rodriguez_X_ray_main_sequence}, and in  accordance with \citeauthor{2024_Rodriguez_X_ray_main_sequence} we find that all of them lie in the region expected for accreting white dwarf systems. Out of 38 transients classified as "CV" in our analysis, 29  had existing \textit{Simbad} classification, 3  were classified as "Star" and 1 was classified as "White Dwarf" in
\textit{Simbad}, while 5 transients didn't have any classification in \textit{Simbad}.

We also found 17 transients, classified as Young Stellar Object or T Tauri Star on {\it Simbad}, to be lying along the main sequence branch in extinction-corrected HR diagram (Figure~\ref{fig:HRD}). We have reclassified these transients and labeled them as "SS/IB".

\item {Classification on the basis of Period information}:
For the $\sim$70\% of the transients that have photometric data in Gaia DR3, we inspected their phase folded lightcurves and power density spectra using the LS periodogram technique.   
Out of these, 130 transients give reasonable phase folded lightcurves, based on our visual inspections.
We report the Periods ($P$), associated with the Gaia lightcurves, for these transients in Table~\ref{Transients table}. 
In order to understand the periodicity associated with different kinds of binary systems and
to facilitate source classification based on the periods, we compiled the periods associated with the different classes of X-ray sources.  We show in Figure~\ref{fig:Period of CV and XRB}, periods associated with some accreting compact object systems like CVs \footnote{ Vizier Table: J/other/NewA/34.234/catalog and J/MNRAS/525/3597/tablea1	 \citep[e.g][]{2015_period_CV,inight_2023_catalogue}}, Low mass X-ray binary systems (LMXBs)\footnote{Vizier Table: J/A+A/684/A124/lmxbcat  \citep[e.g][]{2024_Fortin_period_LMXB}}, High Mass X-ray binary systems (HMXBs) \footnote{Vizier Table: J/A+A/671/A149/tablea  \citep[e.g][]{2023_Fortin_period_HMXB}}, as well as with some non-accreting systems like RS CVn-type systems \footnote{Vizier Table: J/ApJS/249/18/table2 \citep[e.g][]{2020_chen_period_RS_CVn}} and Algol-type systems \footnote{Vizier Table: J/ApJS/238/4/binaries  \citep[e.g][]{2018_period_algols}}.
For the newly identified accreting compact object candidates (discussed in Section~\ref{sec:new detection section}), we show the phase-folded light curves in the appendix.

\begin{enumerate}
   
\item 
We labeled transients as "IB/ACO" (total 50 transients, 3 of which did not have any \textit{Simbad} classification) if their periods (found from optical data) were between a few minutes and $\sim$8 hours, since such (orbital) periods correspond to accreting compact objects or close interacting binaries (Figure~\ref{fig:Period of CV and XRB}; see also \citealt{inight_2023_catalogue}). 
    \item 
41 transients have luminosity class III (taken from {\it Simbad} or derived using the temperature-radius information in Gaia DR2 along with e.g. Appendix G of \citet{1996_book_Carroll_Ostlie}) , having mean $L_X \sim 4\times10^{31}$ erg s$^{-1}$. Out of these 41 transients, one transient is labeled as "LPV" (because of existing \textit{Simbad} class and large orbital period), two are labeled as "CV" (on the basis of their position in HR diagram, large $L_X$ and short periods) and 38 transients have been labeled as "IB" 
 (see also Section~\ref{sec:summary}).
\item 
The remaining 39 transients (out of which, 4 did not have any \textit{Simbad} classification) were labeled as "YSO", "SS/IB", "IB" and "CV" on the basis of their position in HR diagram, Period, $L_X$ and {\it Simbad} Class.
\end{enumerate}

\item Finally, we revisited all source classifications from {\it Simbad} to ensure that they conform with the above criteria. In summary we reclassified $\sim$50 transients that we believe were misclassified or insufficiently classified on {\it Simbad}, and newly classified the $\sim$100 transients that had no class in {\it Simbad}.
\end{enumerate}



\begin{figure}
     \centering
     \includegraphics[width=0.45\textwidth]{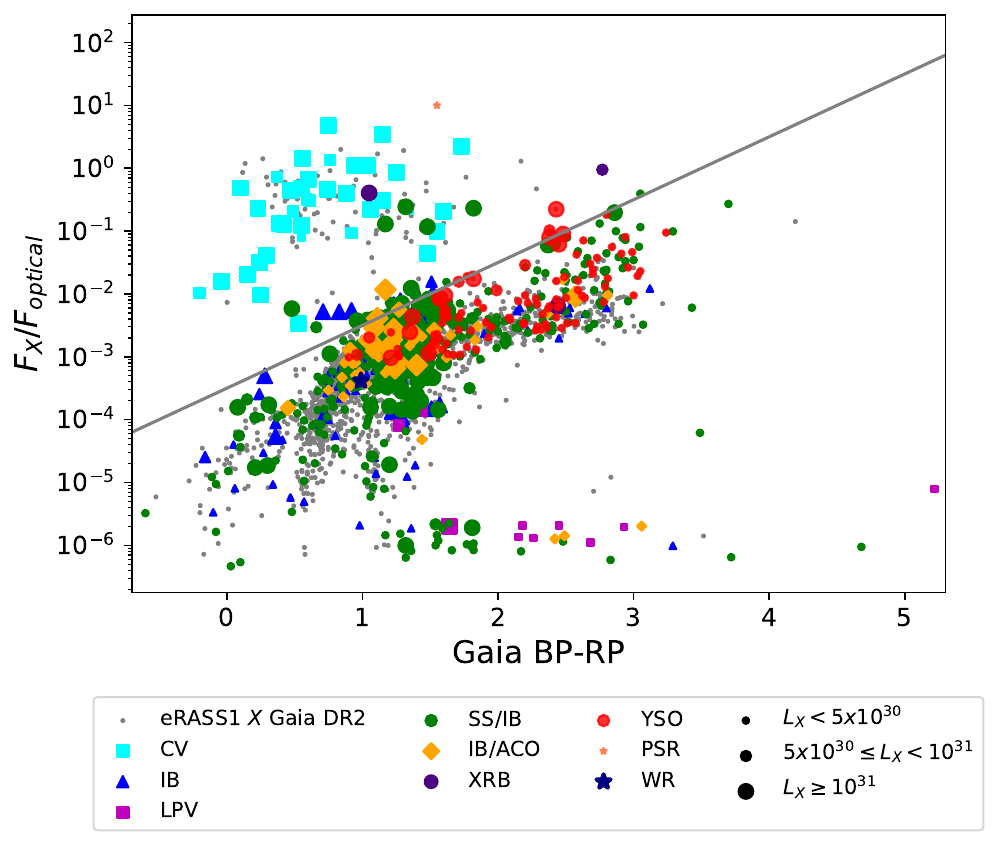}
     \caption{The "X-ray main sequence" plot \citep[c.f.][]{2024_Rodriguez_X_ray_main_sequence} for the Galactic transients in eRASS1. It separates accreting compact objects in upper left corner, from stars and symbiotic systems using an empirical cut, as shown with the grey straight line.  Y axis shows the ratio of X-ray to optical flux ( optical flux calculated from Gaia's green filter magnitude) while the X-axis shows Gaia color. Varying marker correspond to the different X-ray luminosities (in erg s$^{-1}$) of the transients, as shown in the legend.}
     \label{fig:x-ray-main-sequence}
 \end{figure}

\begin{figure} 
    \centering
    
    \includegraphics[width=0.45\textwidth]{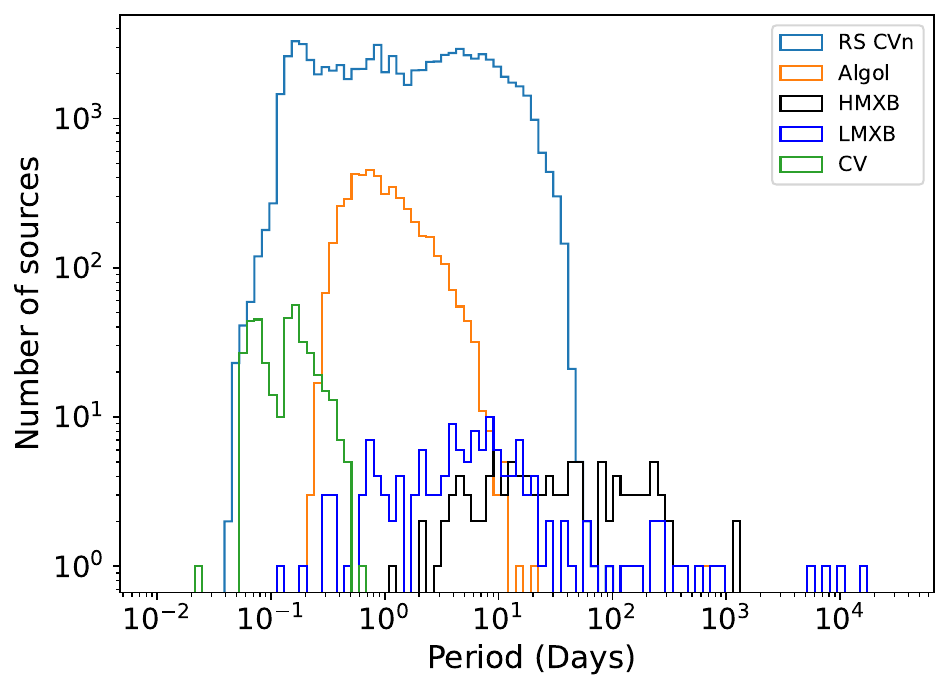} 
    \caption{Compilation of the periods associated with the different classes of X-ray sources, executed to facilitate X-ray transient source classifications based on the periods. Here we show the histogram of orbital or rotation periods associated with CV \citep[e.g][]{2015_period_CV,inight_2023_catalogue}, High Mass X-ray binary (HMXB) \citep[e.g][]{2023_Fortin_period_HMXB} , Low Mass X-ray binary (LMXB) \citep[e.g][]{2024_Fortin_period_LMXB}, RS CVn \citep[e.g][]{2020_chen_period_RS_CVn} and Algol systems \citep[e.g][]{2018_period_algols}. The periods mentioned refer to orbital periods for CV, HMXB, and LMXB.}
    \label{fig:Period of CV and XRB}
\end{figure}


\begin{figure}
     \centering
      \includegraphics[width=0.45\textwidth]{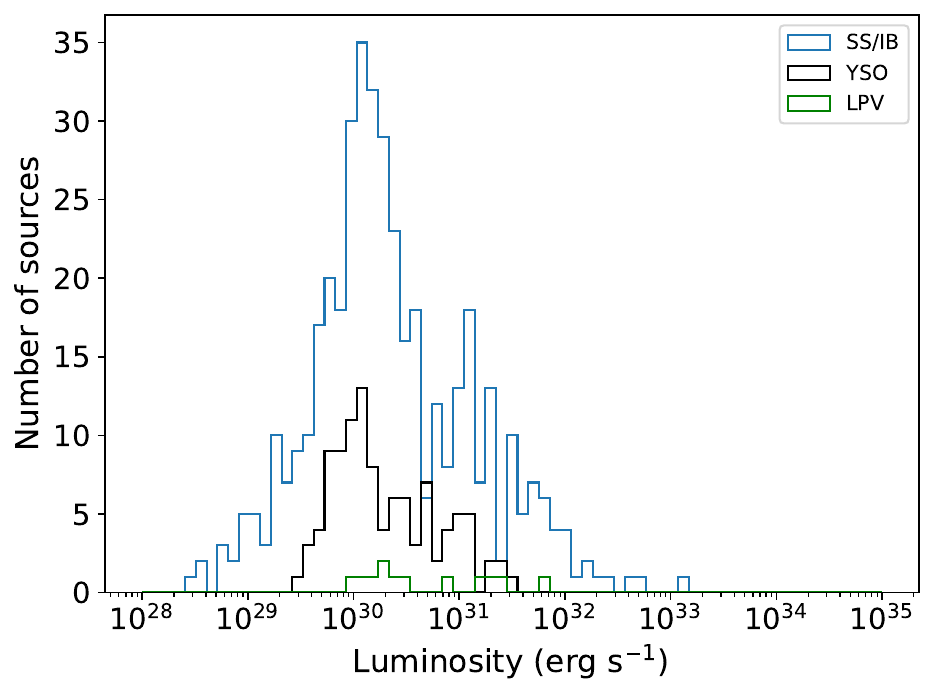}
     \includegraphics[width=0.45\textwidth]{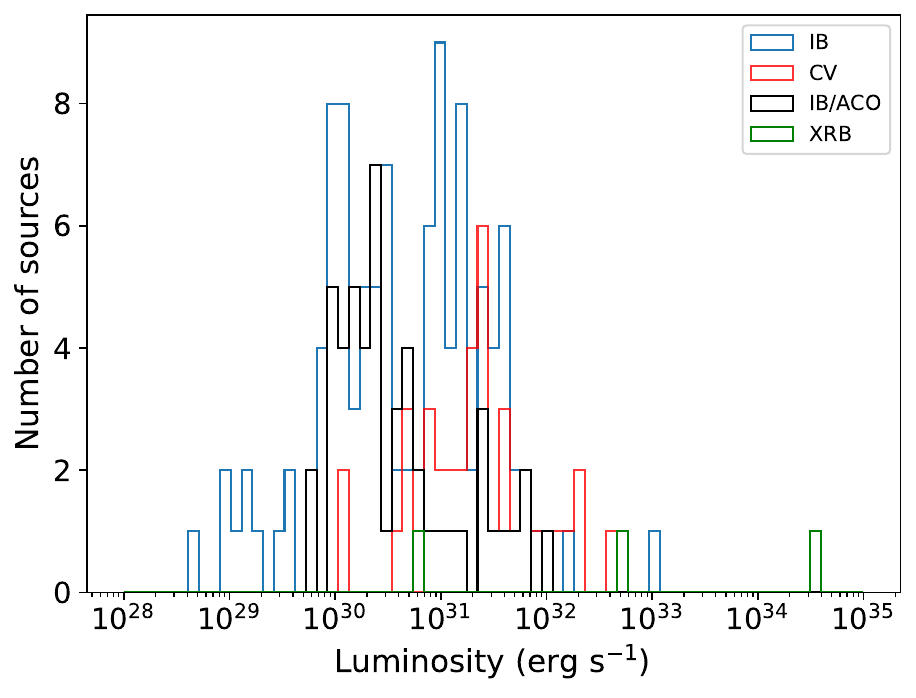}

     \caption{Luminosities of the different classes of X-ray transients found in this work.}
     \label{X-ray luminosity histogram plot of various classes}
 \end{figure}
\begin{figure} 
    \centering
    \includegraphics[width=0.45\textwidth]{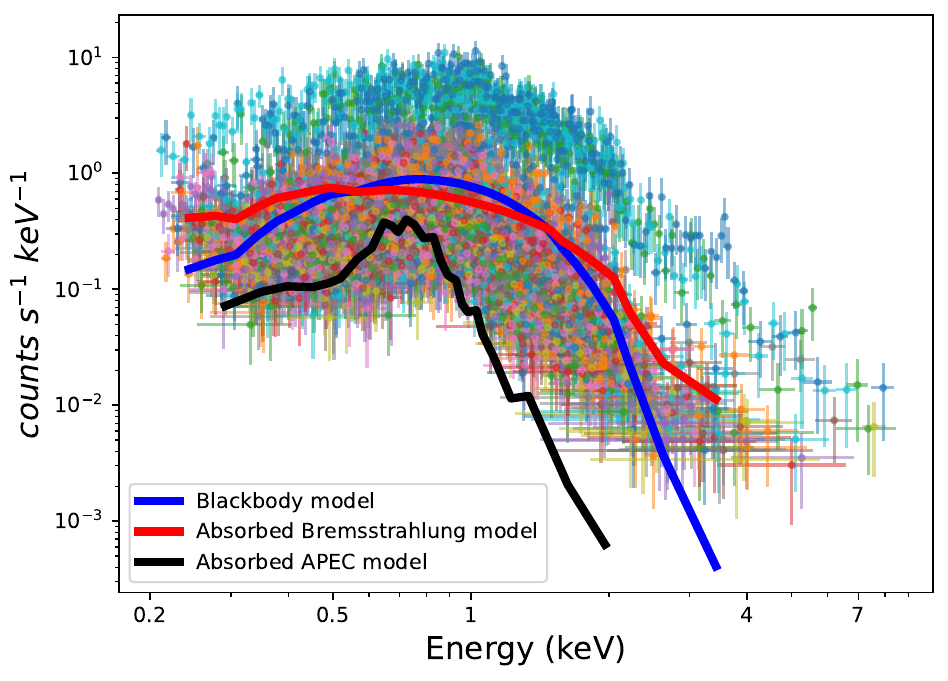} 
    \includegraphics[width=0.45\textwidth]{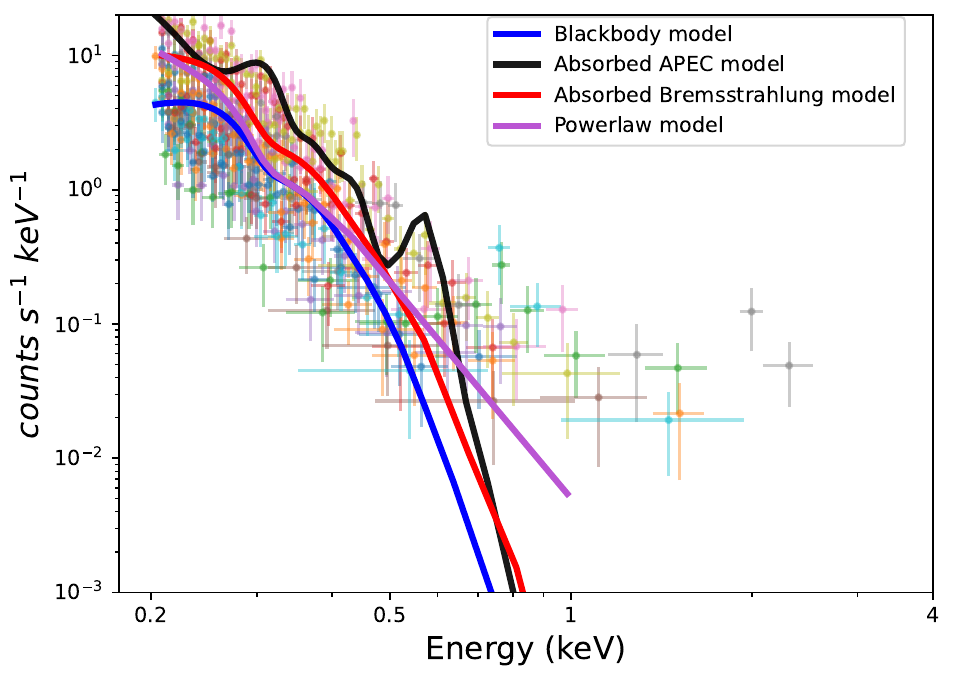}
    \caption{The two distinct types of eROSITA X-ray spectra (spectra peaked around 0.8 keV in the top panel and  monotonically falling spectra shown in the lower panel) found in transients in the "SS/IB", "YSO", "LPV", "IB" and "IB/ACO" categories. 
    Spectral fits using blackbody, thermal bremsstrahlung and APEC models (thick lines, color coded) are overplotted. All the transients in the lower panel are known giant stars. See text for details. }
    \label{fig:X-ray Spectra of B to M spectral type sources}
    
\end{figure}
\begin{figure}
    \centering
    \includegraphics[width=0.45\textwidth,scale=0.4]{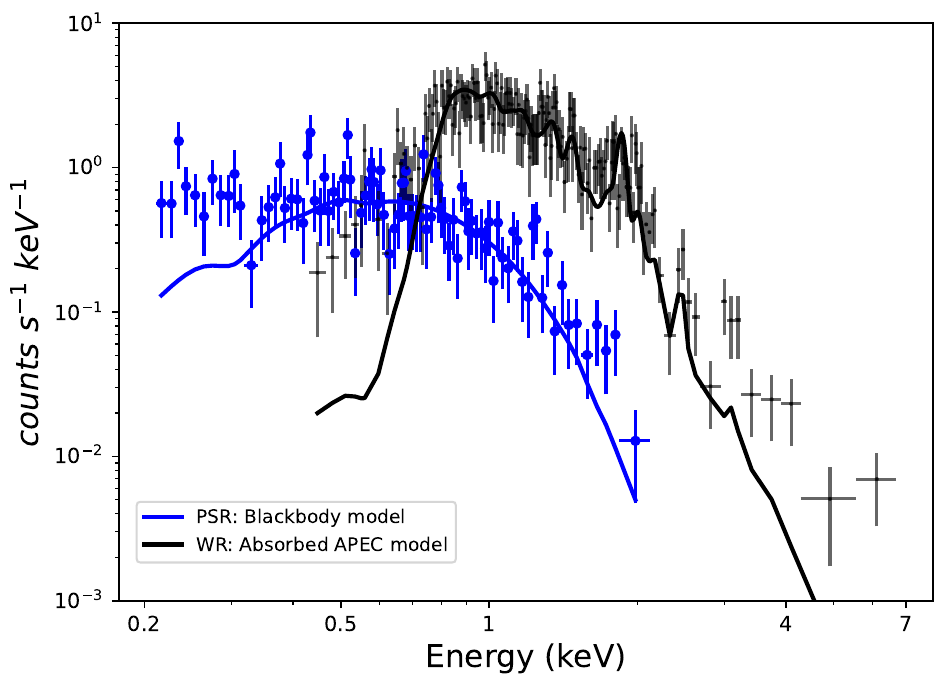}
    \caption{Spectra of transients classified as Wolf Rayet and pulsar. The best-fit models shown as thick curves and marked in the legend. }
    \label{fig:WR AND PSR spectra}
\end{figure}

\begin{figure} 
    \centering
    \includegraphics[width=0.45\textwidth]{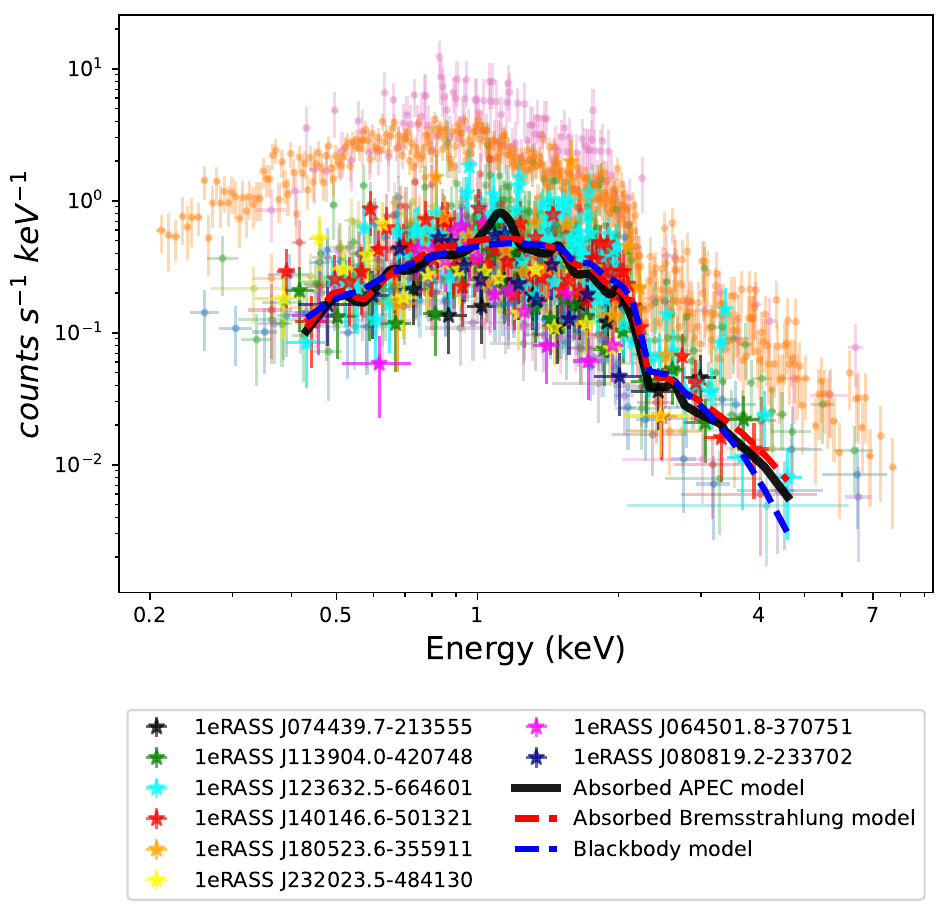} 
    \includegraphics[width=0.45\textwidth]{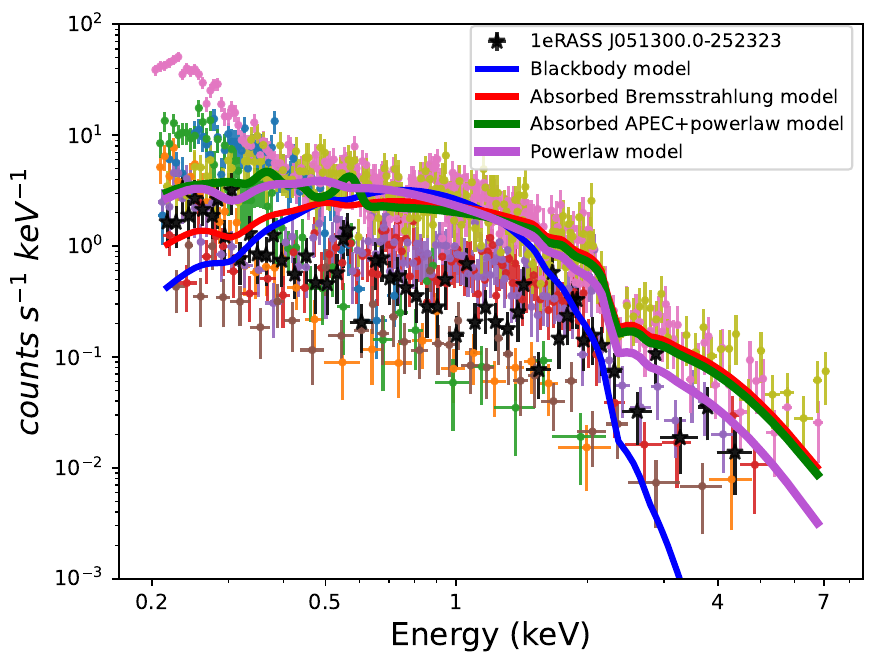} 
    \caption{Spectra of all the known and newly identified  Cataclysmic Variable systems in our sample. 
    The newly identified CVs/candidates are shown with star-shaped markers. \textit{Upper panel}: All CVs showing peaked spectra in the eROSITA band (29 sources), \textit{Lower panels}: CVs having monotonic spectra with spectral features (9 sources) (Their individual unfolded spectra are also shown in the appendix). Spectral fits using blackbody, thermal bremsstrahlung and APEC models (thick lines) are overplotted.}
    \label{fig:CV spectra}
\end{figure}

\begin{figure}
    \centering
    \includegraphics[width=0.45\textwidth,scale=0.4]{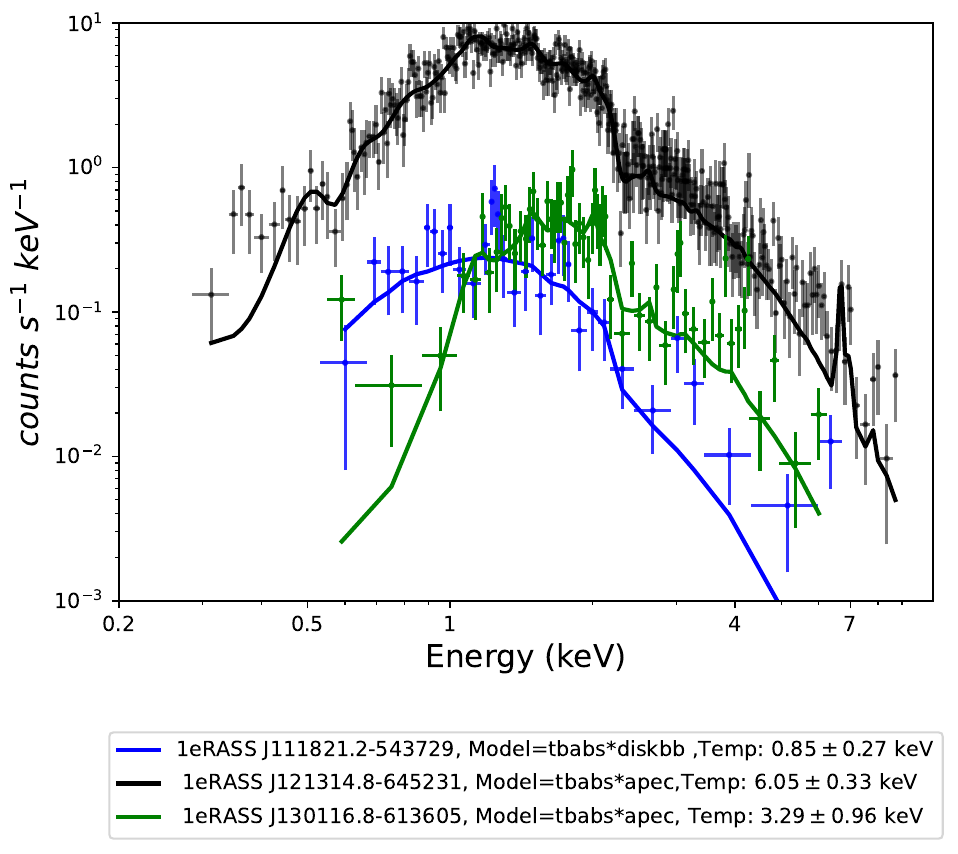}
    \caption{Spectra of X-ray Binaries in our sample of transients, which show extended emission and undulations down to 7-8 keV. Best fit models and associated temperature values are shown in legend.  }
    \label{fig:XRB spectra}
\end{figure}


\section{Properties of the eRASS1 Galactic transients}\label{Property discussion section}
\subsection{X-ray luminosity and spectral properties of different transient classes}

\begin{table}
\centering
\caption{X-ray luminosity ranges and peak luminosities for various transient source classes. These values are derived from the luminosity histograms shown in Figure~\ref{X-ray luminosity histogram plot of various classes}.}
\begin{tabular}{lcc}
\hline
\textbf{Source Class} & 
\makecell{\textbf{Luminosity Range} \\ (erg s$^{-1}$)} & 
\makecell{\textbf{Peak Luminosity} \\ (erg s$^{-1}$)} \\
\hline 
\hline
\rule{0pt}{0.9\normalbaselineskip}CV & $1 \times 10^{30}$ -- $5 \times 10^{32}$ & $\sim 2 \times 10^{31}$ \\
IB & $5 \times 10^{28}$ -- $1 \times 10^{33}$ & $\sim 1 \times 10^{31}$ \\
IB/ACO & $5 \times 10^{29}$ -- $2 \times 10^{32}$ & $\sim 2 \times 10^{30}$ \\
LPV & $1  \times 10^{30}$ -- $1 \times 10^{32}$ & $\sim 2 \times 10^{30}$ \\
PSR &  ----  & $= 1.8 \times 10^{30}$ \\
SS/IB & $2 \times 10^{28}$ -- $1 \times 10^{33}$ & $\sim 2 \times 10^{30}$ \\
WR & ---- & $= 2.5 \times 10^{33}$ \\
XRB & $5 \times 10^{30}$ -- $5 \times 10^{34}$ & ---- \\
YSO & $3 \times 10^{29}$ -- $4 \times 10^{31}$ & $\sim 1 \times 10^{30}$ \\
\hline
\end{tabular}
\end{table}


    Transients that are in "SS/IB", "YSO", "LPV", "IB", "IB/ACO" categories show two distinct kinds of X-ray spectra as shown in Figure~\ref{fig:X-ray Spectra of B to M spectral type sources} .
     Transients in the upper panel exhibit similar spectra, having a shallow rising part between 0.2--0.8 keV, peaks between $\sim$ 0.8--1 keV, and fairly steep decline in count rate beyond 1 keV. The majority of the transients have spectra that are seen to terminate at $\sim$3--4 keV. In contrast, the transients in the lower panel show much softer spectra, peaking below 0.2 keV. Their spectra are steeply declining monotonically, and the majority of the transients spectra are seen to terminate around 2 keV.
     Interestingly, all of these monotonically declining spectra are associated with giant stars.

     We fit most of the spectra using unabsorbed blackbody (\textit{bbody}), absorbed thermal bremsstrahlung (\textit{tbabs*bremss}) and/or  absorbed APEC (\textit{tbabs*apec}; abundance fixed to solar abundance) using \textit{PyXspec}. 
     The model {\it tbabs} takes into account the Galactic absorption \citep[e.g][]{2000_tbabs_model}, while the APEC model accounts for collisionally-ionized optically thin plasma emission \citep[e.g][]{2001_smith_APEC}. 
     Spectra of the transients in the upper panel of Figure~\ref{fig:X-ray Spectra of B to M spectral type sources} give a best fit with a blackbody model of temperature 0.24 keV (mean reduced $\chi^2 \sim 1.05$), while significantly increased values of reduced $\chi^2$ ($\sim1.6$) were obtained for the best fit APEC and bremsstrahlung models where the mean temperature is 0.8 keV.
     Transients in the lower panel of Figure~\ref{fig:X-ray Spectra of B to M spectral type sources} show good spectral fit only with a powerlaw model, having mean powerlaw photon index $\Gamma \sim$ 6.6. 

The X-ray spectra of "WR" and "PSR" transients are shown in Figure~\ref{fig:WR AND PSR spectra}.
The pulsar has a flat spectrum between 0.2--1 keV, and beyond this range it shows a steep decline in counts up to 2 keV. We find the best fit model to be blackbody with a temperature of 0.18 keV. The spectrum of the Wolf Rayet shows steep rise between 0.5--1 keV, followed by sharp decline in counts. The WR spectrum shows best fit with the APEC model (temperature = 0.75 keV).

    
Figure~\ref{fig:CV spectra} shows the eROSITA spectra of CVs, divided into two panels to portray the different spectral shapes. 
CVs showing spectra peaking at $\sim$1--1.5 keV are shown in the top panel while the bottom panel shows spectra peaking at energies $\lesssim$0.2\,keV and monotonically declining in the 0.2 to 5--7 keV range. 
The previously unclassified/incompletely classified transients from \textit{Simbad} (now newly classified as CVs in our work) are shown with bold asterisk markers. 
The X-ray spectra of most of the transients in the upper panel of Figure~\ref{fig:CV spectra} show best fit with the APEC model having the mean temperature of 0.8 keV. 
Transients in the lower panel of Figure~\ref{fig:CV spectra} show good fits with powerlaw model, having mean $\Gamma \sim$ 3.1 and with absorbed APEC+powerlaw model having mean $\Gamma \sim$ 1.6 and mean temperature $\sim$ 0.25 KeV.  

The eROSITA spectra of the three XRBs, detected as transients in our search, are shown in Figure~\ref{fig:XRB spectra}. 
Two of these transients show good fits with both APEC and bremsstrahlung models, but APEC models with temperatures between 0.8--6.0 keV give reduced $\chi^2$ values closer to unity and more precise best-fit parameters. 1eRASS J111821.2-543729 shows best fit with absorbed multi temperature blackbody (tbabs*diskbb) model.

\subsection{Notes on selected compact object systems/candidates and Wolf-Rayet system}\label{sec:new detection section}

Here we describe the properties of newly-identified cataclysmic variables systems/candidates, i.e. X-ray transients whose CV classification is missing or incomplete in the \textit{Simbad} and Vizier databases, together with the properties of X-ray binaries, Wolf-Rayet and pulsar, that were found to be transients. 

\subsubsection{1eRASS J051300.0-252323}
This transient is not detected in the 2RXS catalog and has a flux of $1.50\times10^{-12}$ erg s$^{-1}$ cm$^{-2}$ in eRASS1, implying an increase in flux by a factor of $\gg$4.3 between the two surveys. 
Its eROSITA spectrum (lower panel of Figure~\ref{fig:CV spectra}) shows some emission-like features, at $\sim$0.6\, keV, 1 \,keV\, and\, 1.6\,keV energies, superposed on a monotonic continuum between 0.2--2.4 keV decreasing with energy.  
Its Gaia counterpart is located below the main sequence on the HR diagram (Figure~\ref{fig:HRD}) and  lies at a distance of 623 pc.
We thus find its X-ray luminosity to be $L_X=6.95\times10^{31}$ erg s$^{-1}$. 
No temperature and radius information is available for this transient in Gaia DR2, neither there is any useful optical timeseries data to facilitate a measurement of the orbital period.
Nevertheless, we classify this transients as a "CV" based on its position in the HR diagram, relatively large $L_X$ and the X-ray spectrum which conforms with the CV class (see also previous subsection).


 \subsubsection{1eRASS J074439.7-213555} 
 
This transient is not detected in the 2RXS catalog and has a flux of $5.2\times10^{-13}$ erg s$^{-1}$ cm$^{-2}$ in eRASS1, implying an increase in flux by a factor of $\gg$1.5 between the two surveys. Its eROSITA spectrum is noisy; it appears to be almost flat between 0.6--2 keV followed by a decline and termination of counts at 3 keV (upper panel of Figure~\ref{fig:CV spectra}).
Its Gaia counterpart is a K III-type giant and is located at a distance of 2653 pc, which implies $L_X=4.39\times10^{32}$ erg s$^{-1}$. 
We find a period of 51 min from the Gaia RP and G band photometric light curves, but we are unable to establish how robust this period is, given the shape and low amplitude of implied variations in the phase-folded light curve (see Figure in the appendix). If this period is robust then it likely does not correspond to the orbital period as the typical orbital periods of binary systems involving giant stars is of the order of few days \citep[e.g][]{2022_beck_orbital_period_involving_giant_stars,2023_Fortin_period_HMXB}, but could indicate the spin period of a compact object. 
The relatively high $L_X$ together with the X-ray spectral shape (including the spectral cutoff around 3 keV) are consistent with the observed properties of accreting white dwarf systems. However, such a white-dwarf-KIII giant binary system is difficult to conceive in terms of a stellar evolution scenario. 
If our derived period is not robust, then an interacting binary classification is most appropriate for this X-ray transient.

\subsubsection{1eRASS J113904.0-420748}

This transient is not detected in the 2RXS catalog and has a flux of $6.87\times10^{-13} $ erg s$^{-1}$ cm$^{-2}$ in eRASS1, implying an increase in flux by a factor of $\gg$ 2 between the two surveys.  
It is located at a distance of 702 pc, implying $L_X=4.05\times10^{31}$\,erg s$^{-1}$. 
Its eROSITA spectrum shows a sharp peak around 1 keV (upper panel of Figure~\ref{fig:CV spectra}). It is classified as "Star" in \textit{Simbad}. Its spectral luminosity class is G V, and the period information cannot be ascertained from Gaia DR3.
We classify this transient as "CV" because of its position in HR diagram and relatively large $L_{X}$.

\subsubsection{1eRASS J123632.5-664601 } 

This transient is not detected in the 2RXS catalog and has a flux of $1.45\times10^{-12} $ erg s$^{-1}$ cm$^{-2}$ in eRASS1, implying an increase in flux by a factor of $\gg$ 4.2 between the two surveys. 
Its eROSITA spectrum is peaked around 1 keV with possible spectral features between 0.8-2 keV (upper panel of Figure~\ref{fig:CV spectra}).
Its Gaia optical counterpart is below the main sequence in the HR diagram  and lies at the distance of 360 pc. We thus find its  $L_X=2.24\times10^{31}$erg s$^{-1}$. It is classified as "Star" in \textit{Simbad}. No temperature or radius data is available for the transient in Gaia DR2, and period information is also missing in Gaia DR3. 
Nevertheless, we classify this transient as a "CV" based on its position in the HR diagram and the X-ray spectrum that conforms with the CV class.


\subsubsection{1eRASS J140146.6-501321}
This transient is not detected in the 2RXS catalog and has a flux of $1.19\times10^{-12} $ erg s$^{-1}$ cm$^{-2}$ in eRASS1, implying an increase in flux by a factor of $\gg$ 3.4 between the two surveys. 
It is located at a distance of 1187 pc, and thus $L_X=1.99\times10^{32} $ erg s$^{-1}$. 
The eROSITA spectrum of this transient shows a broad peak between 0.5--2 keV (upper panel of Figure~\ref{fig:CV spectra}).  
Its Gaia optical counterpart is located at the position of typical K-type main sequence stars in the HR diagram, indicating spectral-luminosity class K V.
We find a period of 111 min using the Gaia DR3's G filter photometric data which we consider as orbital period of the system. 
We classify this transient as a "CV" because of its relatively large $L_X$, as compared to typical K-type main sequence stars, and relatively short period ( Typical orbital periods of CVs are shown in Figure~\ref{fig:Period of CV and XRB})



\subsubsection{1eRASS J180523.6-355911}
This transient is not detected in the 2RXS catalog and has a flux of $1.26\times10^{-12} $ erg s$^{-1}$ cm$^{-2}$ in eRASS1, implying a increase in flux by a factor of $\gg$ 3.6 between the two surveys. It is located at a distance of 442 pc, implying $L_X=2.94\times10^{31} $ erg s$^{-1}$. 
The eROSITA spectrum is noisy with potentially a peak around 1.5 keV (upper panel of Figure~\ref{fig:CV spectra}). 
Its Gaia optical counterpart is below the main sequence in the HR diagram, has temperature of 6858 K. 
Optical light curve data are sparse so the value of period cannot be ascertained.
We classify the transient as "CV" because of its position in HR diagram and relatively large $L_{X}$.


 \subsubsection{1eRASS J232023.5-484130}
This transient is not detected in the 2RXS catalog and has a flux of $7.93\times10^{-13} $ erg s$^{-1}$ cm$^{-2}$ in eRASS1, implying an increase in flux by a factor of $\gg$ 2.3 between the two surveys.  
It is located at a distance of 475 pc, implying $L_X=2.14\times10^{31}$ erg s$^{-1}$. 
The transient has a flat spectrum between 0.4--1 keV followed by a decline and counts terminating at 2.5 keV (upper panel of Figure~\ref{fig:CV spectra}). 
Its Gaia optical counterpart is below the main sequence in the HR diagram. 
There is no temperature and radius information available in Gaia DR2, and the period information cannot be ascertained from Gaia DR3.
We classify this transient as "CV" because of its position in HR diagram and relatively large $L_{X}$.

\subsubsection{1eRASS J064501.8-370751}
This transient is detected in the 2RXS catalog as 2RXS J064502.3-370753 and has a flux of  $5.82\times10^{-13} $ erg s$^{-1}$ cm$^{-2}$ in eRASS1; the increase in flux is by a factor of 4.5 between the two surveys.
It is located at a distance of 523 pc, implying $L_X=1.90\times10^{31}$ erg s$^{-1}$.
Its eROSITA spectrum shows a sharp peak around 0.8 keV (upper panel of Figure~\ref{fig:CV spectra}).
Its Gaia optical counterpart is below the main sequence in the HR diagram.
While earlier this source did not have any known classification on \textit{Simbad}, at the time of writing of this paper, the \textit{Simbad} classification is "white dwarf" \citep[e.g][]{2024_Vincent_New_white_dwarf_in_gaia_era}.
No temperature and radius information of the source is available in  Gaia DR2. 
We obtain a period of 50 min from Gaia DR3's G filter photometric timeseries data, which is consistent with the orbital periods of CVs (Figure~\ref{fig:Period of CV and XRB}) -- specifically the AM CVn-type (having orbital period in the range of 5 to 65 minutes;  \citealt{2018_Ramsay_AM_CVn_Periods}).
Based on these information, we classify this transient as a "CV" of AM CVn type.

\subsubsection{1eRASS J080819.2-233702}
This transient is detected in the 2RXS catalog as 2RXS J080820.0-233652 and has a flux of $8.3\times10^{-13} $ erg s$^{-1}$ cm$^{-2}$ in eRASS1; the increase in flux is by a factor of 3.6 between the two surveys.
It is located at a distance of 666 pc, implying  $L_X=4.4\times10^{31}$ erg s$^{-1}$.
Its eROSITA spectrum shows a sharp peak around 1 keV (upper panel of Figure~\ref{fig:CV spectra}).
Its Gaia optical counterpart lies below the main sequence in the HR diagram. It is classified as "Star" in \textit{Simbad}. 
Its spectral luminosity class is K III, and the period information cannot be ascertained from Gaia DR3.
We classify this transient as "CV" based on its position in the HR diagram and the relatively large $L_{X}$.

\subsubsection{1eRASS J104410.2-594310: Wolf-Rayet star WR 25}
This is a well-studied system in literature consisting of an O-type star and a Wolf Rayet star (WN6h+O4f) \citep[e.g.][]{2001_WR25_HUTCH}. 
The X-ray transient is not detected in the 2RXS and has a flux of $4.88\times10^{-12} $ erg s$^{-1}$ cm$^{-2}$ in eRASS1; the increase in flux is therefore a factor of $\gg$14 between the two surveys.
It is located at a distance of 2099 pc, implying $L_X=2.56\times10^{33}$ erg s$^{-1}$. 
We were not able to ascertain the period  associated with this system from the Gaia DR3 timeseries data. 

The eROSITA spectrum shows a broad peak around 1 keV and it is best fit with an APEC model having plasma temperature of 0.75 keV, which appears to be consistent with a scenario where time-varying shocks are generated in colliding winds  \citep[e.g.][]{1987_Pollock_X_ray_emission_from_colliding_winds_in_WR_systems}.

\subsubsection{1eRASS J043715.9-471509: PSR J0437-4715}
PSR J0437-4715 / 4FGL J0437.2-4715 is a well known millisecond accreting pulsar also detected in gamma-rays. The X-ray transient is not detected in the 2RXS and has a flux of $1.06\times10^{-12} $ erg s$^{-1}$ cm$^{-2}$ in eRASS1; the increase in flux is therefore by a factor of $\gg$3 between the two surveys.
It is located at a distance of 120 pc, implying $L_X=1.82\times10^{30}$ erg s$^{-1}$.
We were unable to ascertain the period associated with this system from the Gaia DR3 timeseries data.  
Its eROSITA X-ray spectrum shows a good fit with a blackbody model with a temperature of 0.18 keV.
The thermal origin of the X-ray transient emission may be associated with the neutron star surface 
and/or from an accretion disk.

\subsubsection{X-ray binaries: 1eRASS J111821.2-543729, 1eRASS J121314.8-645231, 1eRASS J130116.8-613605}

These 3 X-ray transients are not detected in 2RXS, and indicate a flux change by factors between (at least) $2-50$ across the two surveys. 
They have $5\times10^{30}\lesssim L_X\lesssim5\times10^{34}$. 
We were unable to ascertain the periods associated with these systems from the Gaia DR3 timeseries. 
The X-ray spectra of 1eRASS J121314.8-645231 and 1eRASS J130116.8-613605 show good fits with the APEC models (although thermal Bremsstrahlung gives reasonable fits as well), while X-ray sepctra of 1eRASS J111821.2-543729 show good fit with absorbed multicolor blackbody model. 
The corresponding best-fit temperature values are between 0.8--6 keV (see Figure~\ref{fig:XRB spectra}). 
Two of these transients are known neutron star XRBs while the nature of compact object in 1eRASS J111821.2-543729 is uncertain. 
The optically thin (transient) emission from these systems may be from boundary layer or from the accretion columns (in case of highly magnetized neutron star systems), while the optically thick emission may be from accretion disk or from neutron star surface.

\section{Summary and Discussion}\label{sec:summary}
\begin{figure} 
    \centering
    
    \includegraphics[width=0.45\textwidth]{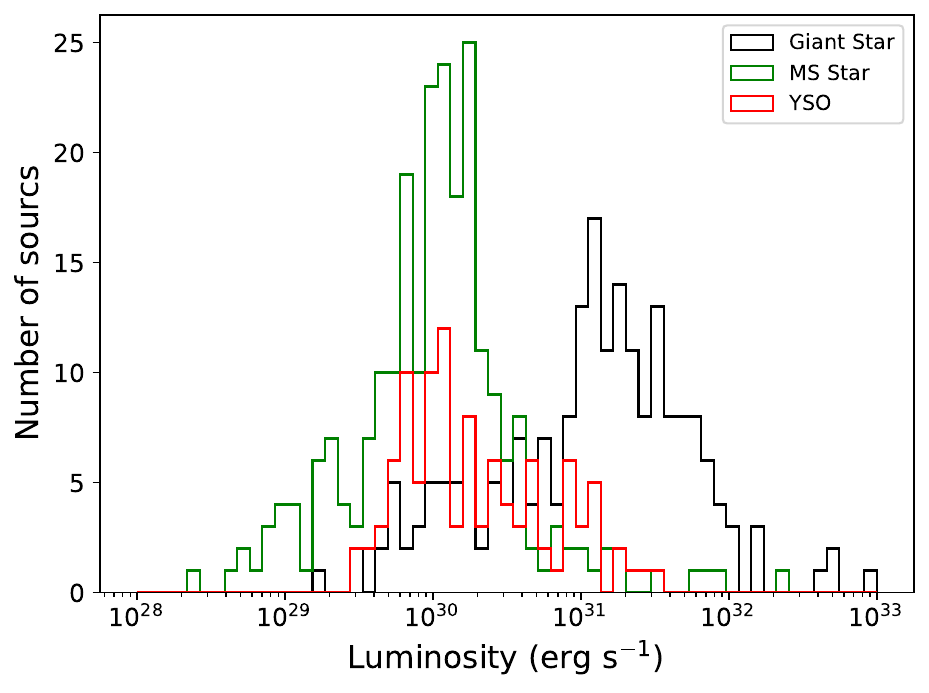} 
    \caption{Luminosity histogram plot of YSOs, main sequence stars and giants in our sample 
    of Galactic transients in eRASS1. Main sequence stars and giants are selected based on their position in the HR diagram.}
    \label{fig:Histogram luminosity plot of Transients and all YSO MSS Gaints,transients only}
\end{figure}

\begin{figure}
    \centering
    \includegraphics[width=0.45\textwidth]{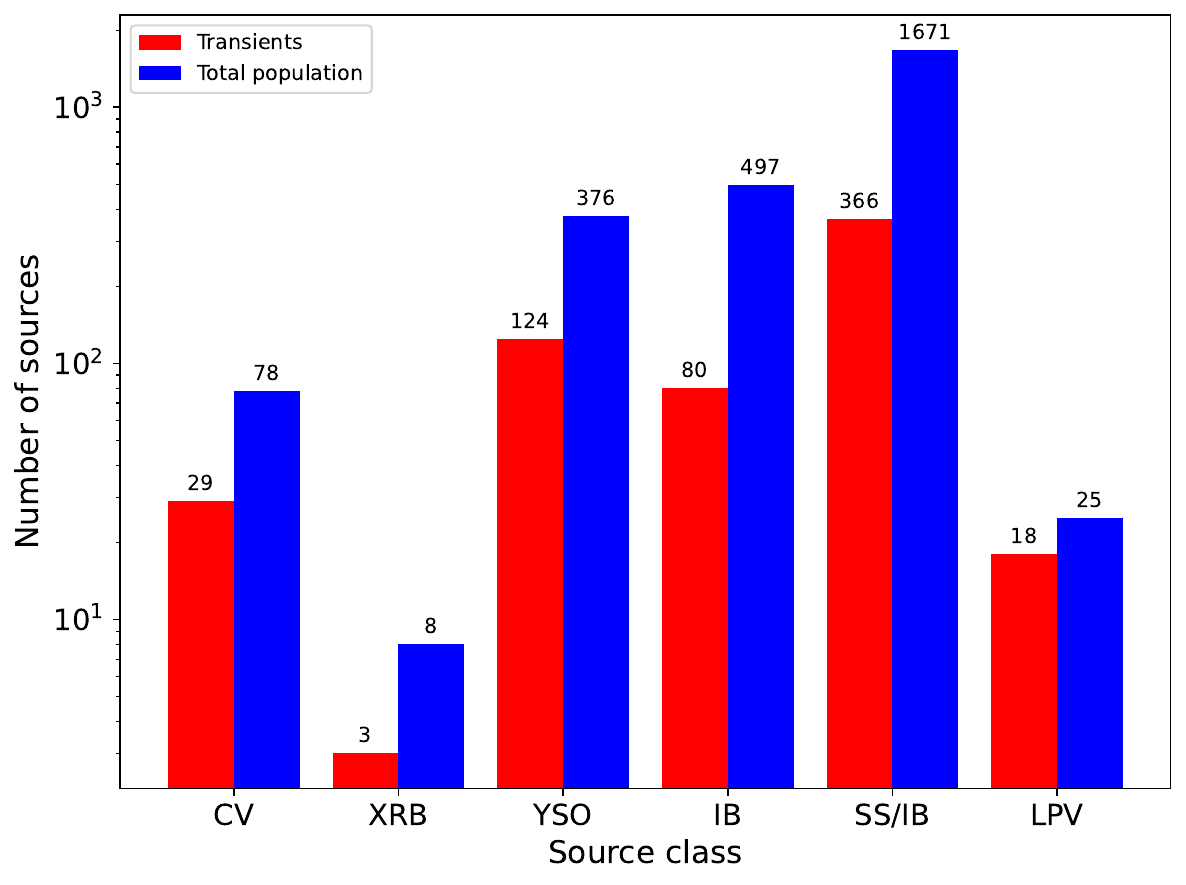}
    \caption{Transient population (red bars) in various classes, considered in this work, compared with the the complete population of X-ray sources (blue bars) found in this work. Each bar has a number above it denoting the total number of sources in that class. The blue bars represent the sums of transient and non-variable (the not variable branch of the flowchart in  Figure~\ref{fig:eROSITA_flowchart}) X-ray sources detected in eRASS1. 
    We have only used the subset of X-ray sources (having counterparts in Gaia) that are accompanied by definite $\textit{Simbad}$ classifications.}
    \label{fig:Histogram_of_different_classes}
\end{figure}

In this paper, we have presented a search for Galactic X-ray transients, having Gaia counterparts, between the eROSITA All Sky Survey 1 (eRASS1) and the Second \textit{ROSAT} All-Sky Survey Point Source (2RXS) catalogs.
This search enables us to find sources that are varying on timescales of $<$30 years. 
We found 738 transients and classified them based on their existing $\it{Simbad}$ classification, their position in the HR diagram, X-ray spectra, optical photometric data from Gaia DR3 and $L_X$ into different categories such as "SS/IB", "IB", "IB/ACO", "LPV", "XRB", "CV", "YSO", "PSR", and "WR". 

We analyzed eROSITA spectral data (0.2--10 keV) for the Galactic transients and fit thermal and non-thermal models, as shown in Figures~\ref{fig:X-ray Spectra of B to M spectral type sources} to~\ref{fig:XRB spectra}.
We found at least two distinct types of spectra for various classes of transients (Figures~\ref{fig:X-ray Spectra of B to M spectral type sources} and ~\ref{fig:CV spectra}).
The first type (upper panels of Figures~\ref{fig:X-ray Spectra of B to M spectral type sources} and ~\ref{fig:CV spectra}) show a peak around 1 keV and are well fit by thermal emission models. 
The second type (lower panels of Figures~\ref{fig:X-ray Spectra of B to M spectral type sources} and ~\ref{fig:CV spectra}) are monotonically decreasing spectra in the eROSITA band (i.e. they peak below 0.2 keV) and require a power-law component (with or without a thermal component) for achieving reasonable fits (31 transients).

It is possible that the monotonic X-ray spectra, requiring a power law component in their fits, are related to magnetic fields in the system. 
For example, magnetic CVs are known to have  soft excess due to reprocessed blackbody radiation, that may give rise to powerlaw-like spectra \cite[e.g.][]{universe_in_x_rays}. 
Moreover, we find that CVs in the upper panel of Figure~\ref{fig:CV spectra} have X ray-to-optical flux ratio between 0.01--1, while those in the lower panel have X ray-to-optical flux ratio in the range 0.1--10. 
The former is characteristic of non-magnetic CVs while the latter is seen in their magnetic analogs \citep[e.g.][]{universe_in_x_rays,2003_mukai_Two_types_of_spectra_in_CVs}. 




In order to investigate if the X-ray luminosities of transients show any systematic trends across the stellar evolutionary sequence, we plotted the X-ray luminosity histograms of our transients that have YSOs, main sequence stars and evolved stars (giants) (see Figure~\ref{fig:Histogram luminosity plot of Transients and all YSO MSS Gaints,transients only}). 
Main-sequence and giant stars have been selected through careful manual inspection of the positions of Gaia counterparts of the transients in the HR diagram (Figure~\ref{fig:HRD}). 
Some previous studies \citep[e.g][]{1972_skumanich_rotation_rate_and_magnetic_activity_relations_in_MS_stars,1999_Feigelson_YSO_More_luminos_then_X_rays_2,1993_Feigelson_YSO_more_luminous_then_MS_stars} have indicated that YSOs are significantly more X-ray luminous compared to main sequence stars, primarily due to their rapid rotation and intense magnetic activity. However, we do not see any such clear trend in our sample of transients. We find the X-ray luminosity distributions to have mean and standard deviations of $30\pm0.6$ erg s$^{-1}$ on log scale for main sequence stars and $30.2\pm0.5$ erg s$^{-1}$ for YSOs. (Figure~\ref{fig:Histogram luminosity plot of Transients and all YSO MSS Gaints,transients only}). 
One other striking feature, however, is that X-ray transients associated with giants are significantly more luminous than those associated only with main-sequence stars or YSOs. 
In a standard scenario we expect stars, as they evolve into giants, to lose angular momentum via stellar flares and mass ejection episodes.
This leads to increased rotation period and decreased magnetic activity, further leading to the expectation that giants should not be very active in X-rays. Some previous studies using \textit{ROSAT} and Einstein observatory have also confirmed that single giants are not very bright in X-rays (specially beyond spectral type K3) \citep[e.g][]{1991_Haisch_Giants_not_active_in_X_rays_ROSAT_observations,Haisch_1992_Giants_not_very_active_in_X_rays,Hunsch_1998_Gaints_not_very_bright_in_x_rays} 
This leads us to speculate that the giants in our transient sample are either part of interacting binary systems or have anomalous magnetic fields associated with them. 

In Figure~\ref{fig:Histogram_of_different_classes} we show a bar plot, for various classes of Galactic X-ray transients found in our work, comparing the number of transients with the total population of Galactic eRASS1 sources (sum of transient and non-variable sources). 
To ensure reliable classification of X-ray sources we have considered only those subset of sources (transients as well as non-variables) that have definite classifications in $\textit{Simbad}$.
Based on this Figure we can find the fraction of sources in any given source class that is expected to show transient behavior  (during a blind observation of the X-ray sky). 
We therefore conclude that 37\% of all CVs, 38\% of all XRB, 33\% of all YSOs, $\sim$20\% of all active stars or interacting binaries and 72\% of all LPVs show transient behavior at any point of time (when compared with their X-ray activity a few decades ago).

Finally, based on the number of transients detected in each source class (see Section 3.1), we calculate the instantaneous all-sky Galactic X-ray transient occurrence rates in a flux limited all sky survey ($F > F_{\rm lim,2RXS}$) as follows. We first consider the completeness factors associated with our catalog cross-matching exercise (see Section 2). Since we are considering sources well beyond the flux thresholds where the completeness drops substantially (see Figure~\ref{fig:flux-hist}), we do not consider the completeness factors associated with the eRASS1 and 2RXS catalogs. In order to investigate the Gaia completeness factors, which we expect to be source class-dependent owing to their different optical-to-X-ray flux ratios, we cross-matched the 20655 eRASS1 sources (having $F_{\rm eRASS1} > F_{\rm lim,2RXS}$) with dedicated catalogs. 
For example, the cross-match of the eRASS1 sources with the Galactic CV catalog 
(\citet{2023_Canbay_CV_catalog_rate_calculation}; matching radius is 5'') resulted in 165 common sources in both catalogs, out of which 79 have counterparts in Gaia DR2 (satisfying the same selection criteria as those given in Section 2.4). Hence, we estimate the Gaia completeness factor for CVs to be $\simeq$2, and the transients occurrence rate is 160 sky$^{-1}$ ($\sim$80 CVs in 2$\pi$ steradian). 
Similarly we find $\sim60$ sky$^{-1}$ for X-ray binaries (using the  \citet{2024_Fortin_period_LMXB} LMXRB catalog, and \citet{2023_Fortin_period_HMXB} HMXRB catalog), and $\sim20$ sky$^{-1}$ for both the pulsar and Wolf-Rayet classes (using the ATNF pulsar catalog \citet{2005_Manchester_ATNF_pulsar_catalog_rate_calculation} and Wolf-Rayet catalog \citet{2020_Rate_Galactic_Wolf_Rayet_catalog_rate_calculation}). 
While we generally expect chromospherically active stars (single stars and interacting binaries in our categories "SS/IB", "YSO", "LPV", and "IB" taken together (645 transients)) to be optically "bright" and have close to 100\% Gaia counterparts, we take a Gaia completeness factor $x$ to find that the rate of these classes as $\sim x \times 1300$  sky$^{-1}$.

\section{ACKNOWLEDGMENTS}
We thank the referee for the valuable comments, which improved the manuscript significantly. We acknowledge the very extensive use of $\textit{Simbad}$, Vizier and X-match services from CDS, Strasbourg, France. This work is based on the data from eROSITA telescope abroad the SRG satellite which is a joint mission between Germany and Russia, \textit{ROSAT} telescope jointly built by Germany, US and UK and the Gaia telescope built by European space agency (ESA).

\section{DATA AVAILABILITY}
eROSITA catalogues and spectral data products are available on the
website \url{https://erosita.mpe.mpg.de/dr1/}, while detailed Table~2 is
available in electronic form at CDS via anonymous ftp to
\url{cdsarc.u-strasbg.fr (130.79.128.5)} or via
\url{https://cdsarc.cds.unistra.fr/viz-bin/cat/J/MNRAS/544/885}.





\bibliographystyle{mnras}
\bibliography{Reference} 
\section{Appendix}
Here we show the periodograms and the phase folded light curves associated with the three transients from Section 4.2, whose period is reported. LS periodograms are noisy but the reported periods are only for those transients, whose phase folded light curves look convincing by visual inspection. We also show the individual eROSITA spectra of the transients mentioned in the lower panel of Figure~\ref{fig:CV spectra}.
\begin{figure*} 
    \centering
    \includegraphics[width=0.45\textwidth]{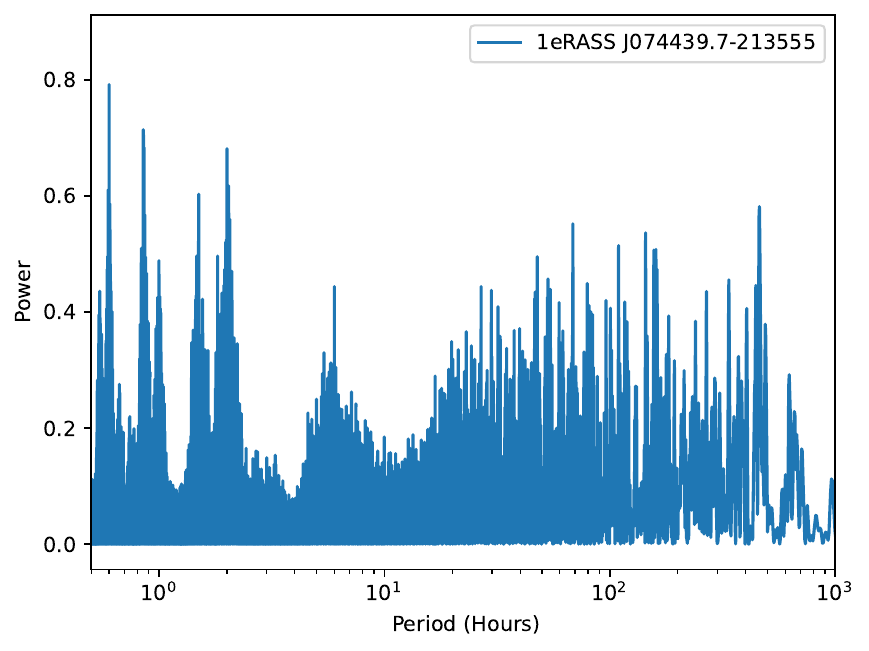} 
    \includegraphics[width=0.475\textwidth]{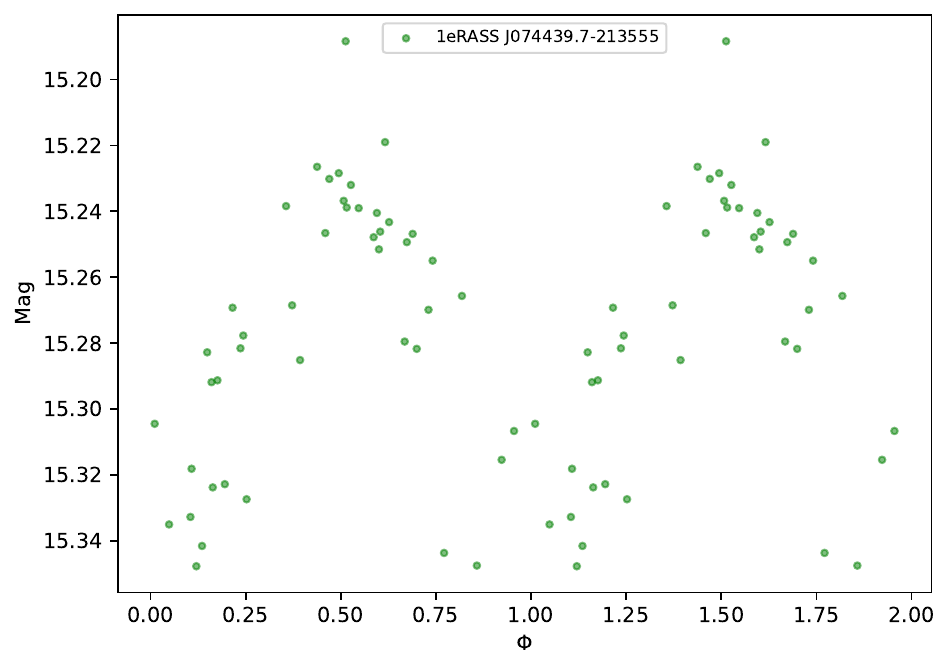} 
    \includegraphics[width=0.45\textwidth]{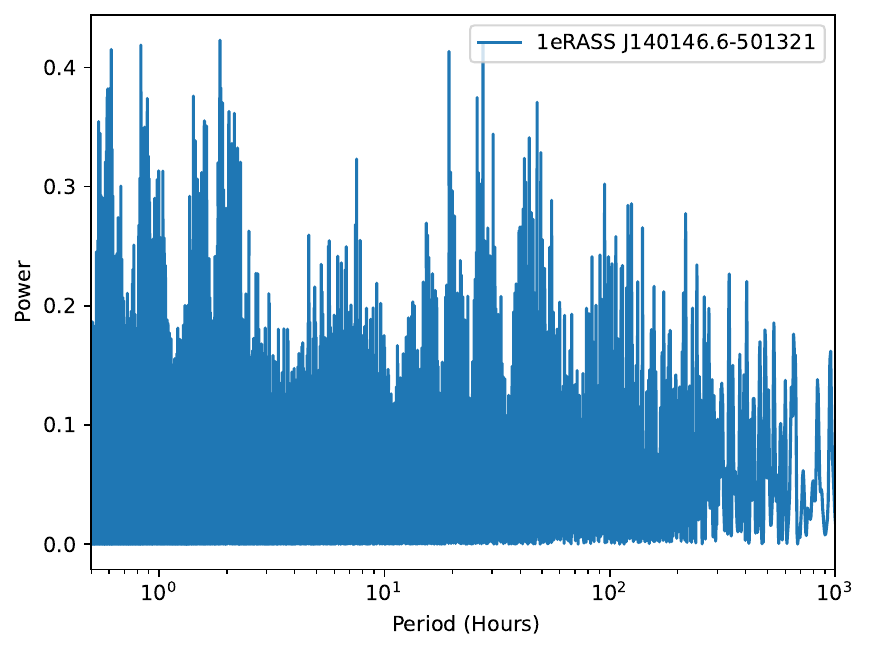}
    \includegraphics[width=0.475\textwidth]{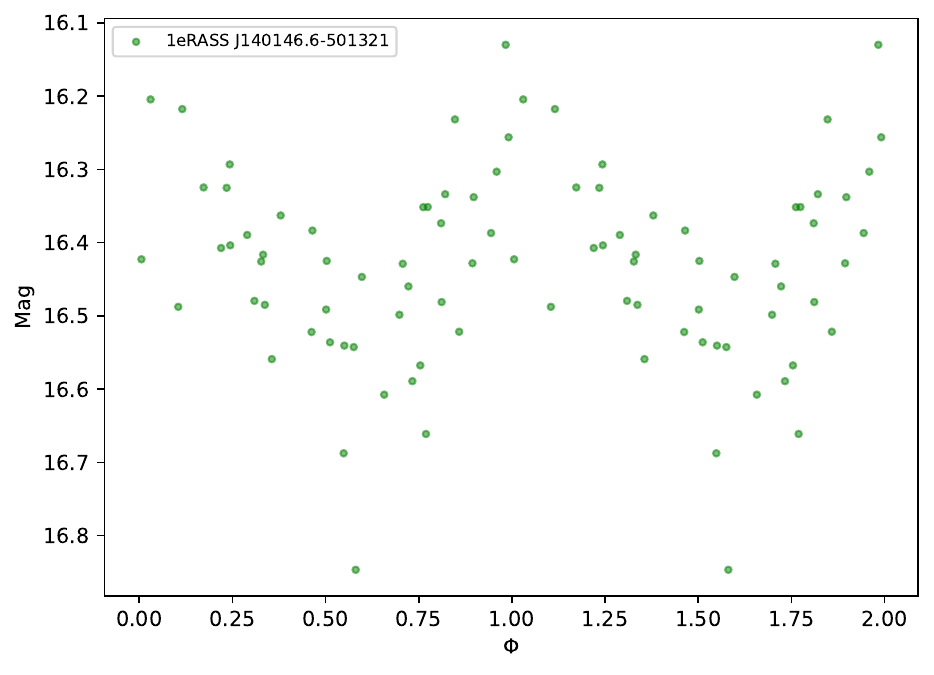} 
     \includegraphics[width=0.45\textwidth]{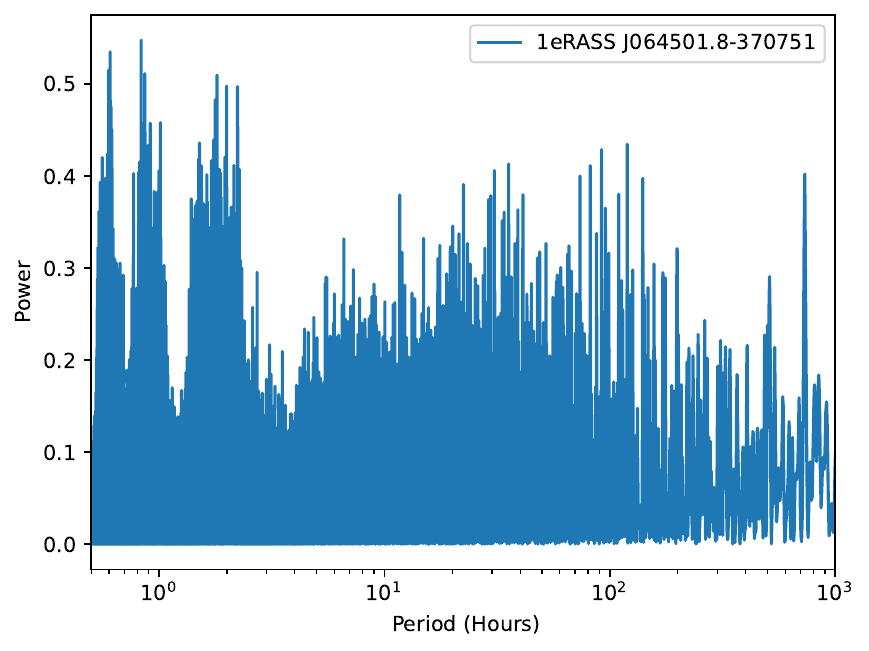}
      \includegraphics[width=0.475\textwidth]{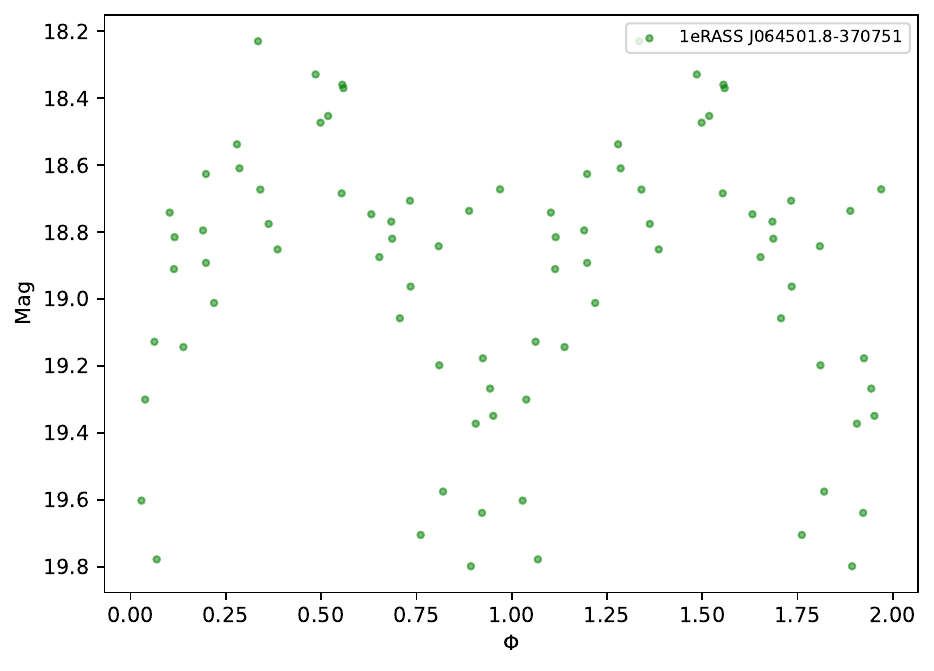}
    \caption{ Periodograms and phase folded light curves associated with the 3 of the 9 new potential CVs identified in this analysis, created from G band photometric data of Gaia DR3. LS periodograms are noisy but the reported periods are only for those transients, whose phase folded light curves look convincing by visual inspection.}
    \label{fig:Phase folded lightcurve}
\end{figure*}

\begin{figure*} 
    \centering
    \includegraphics[width=0.4\textwidth]{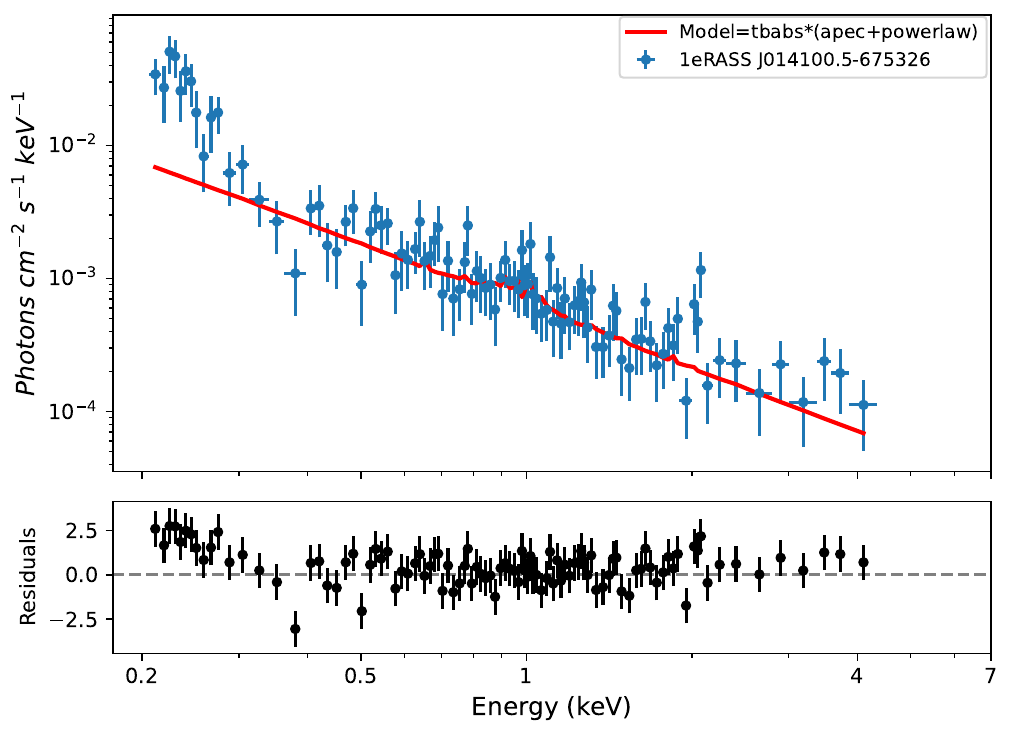} 
    \includegraphics[width=0.4\textwidth]{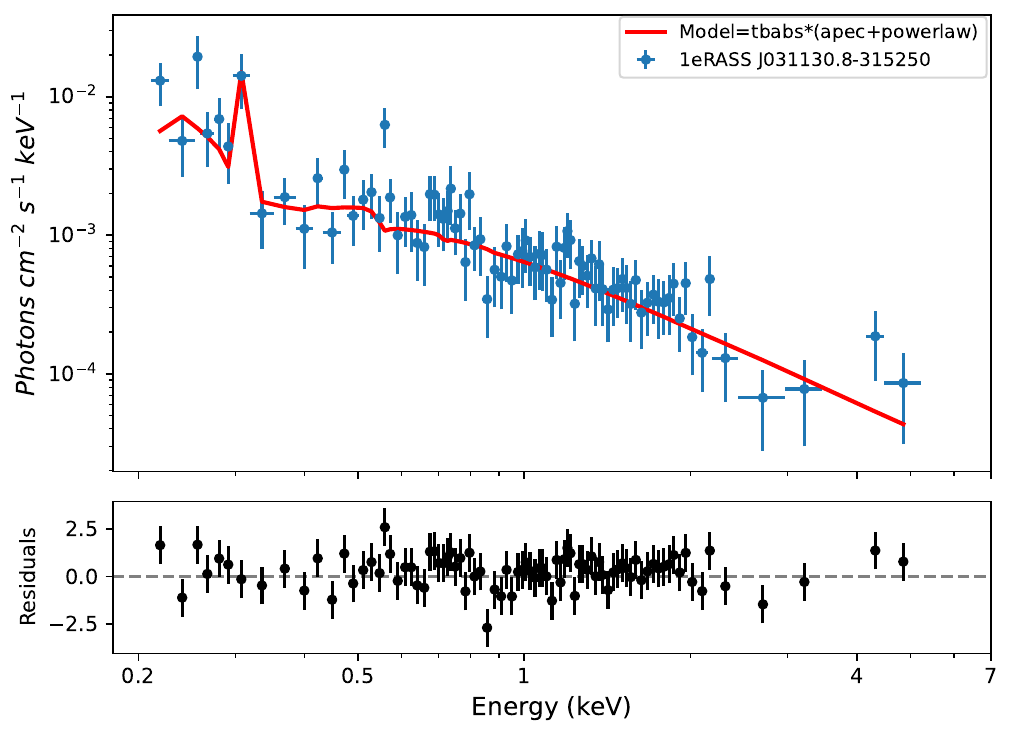} 
    \includegraphics[width=0.4\textwidth]{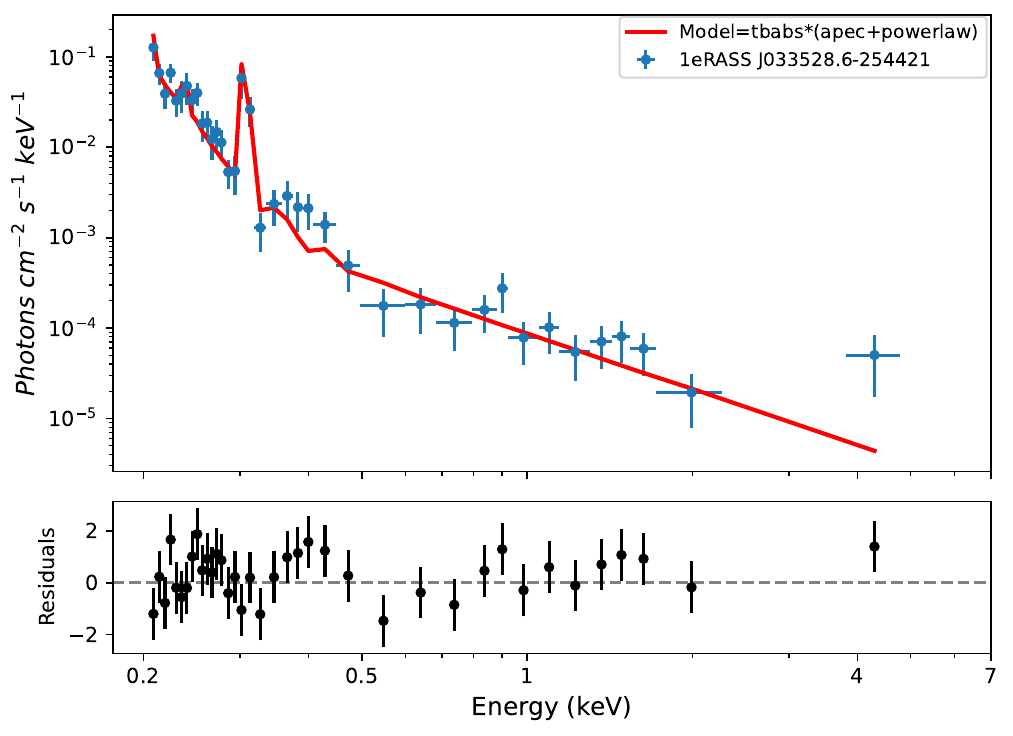}
    \includegraphics[width=0.4\textwidth]{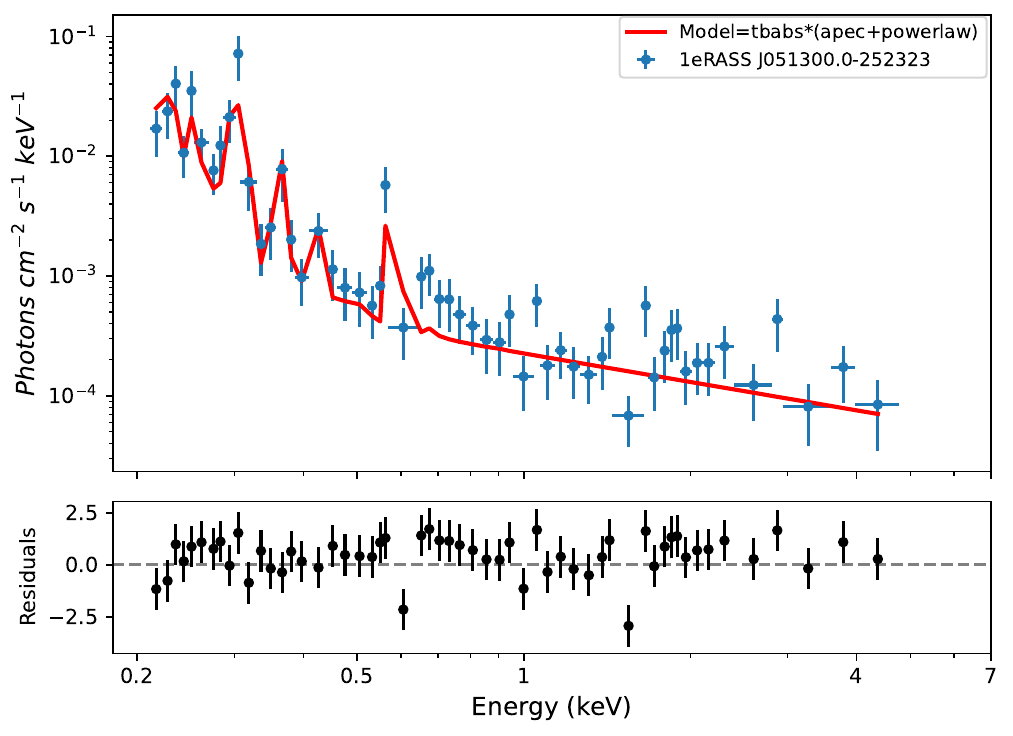} 
     
    \includegraphics[width=0.4\textwidth]{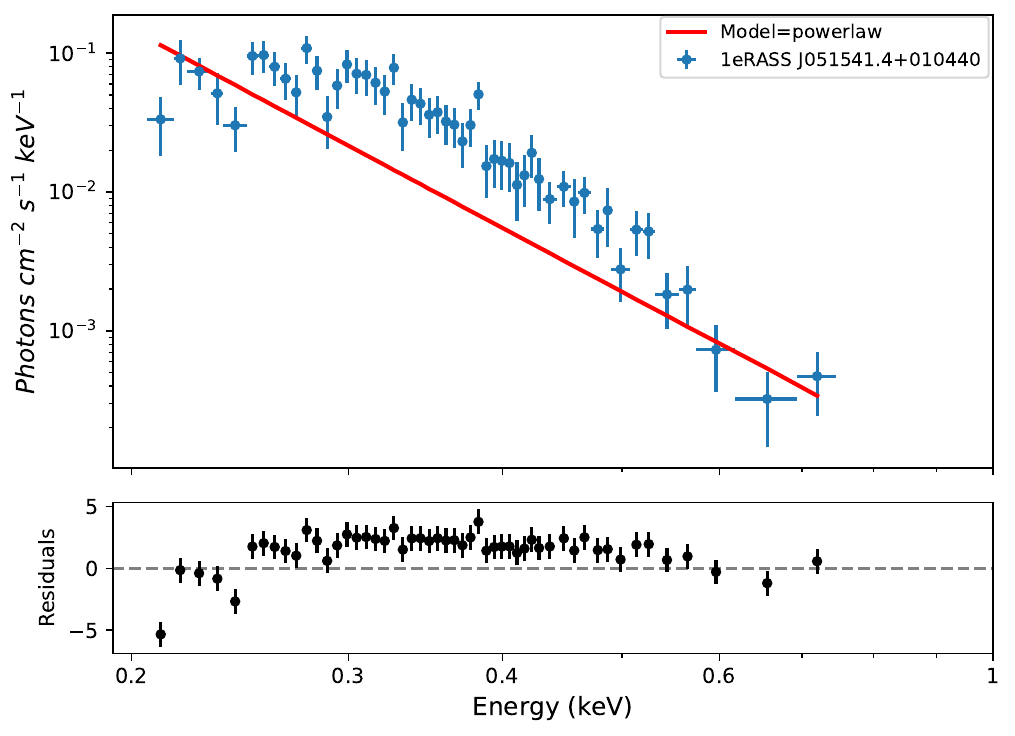}
      \includegraphics[width=0.4\textwidth]{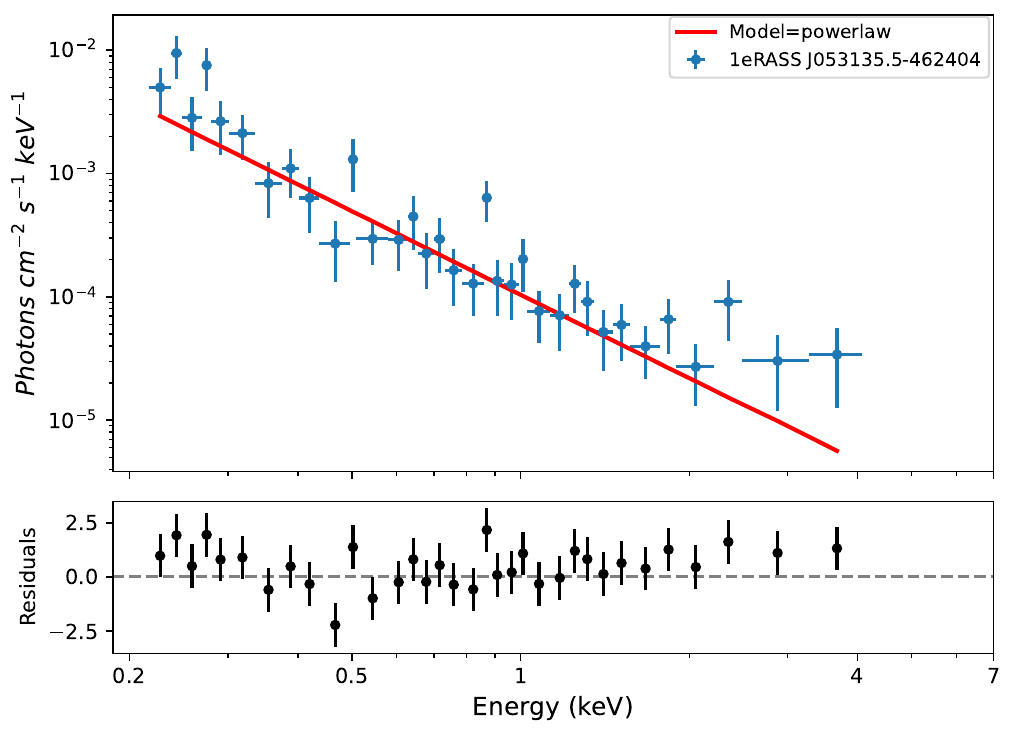}
      \includegraphics[width=0.4\textwidth]{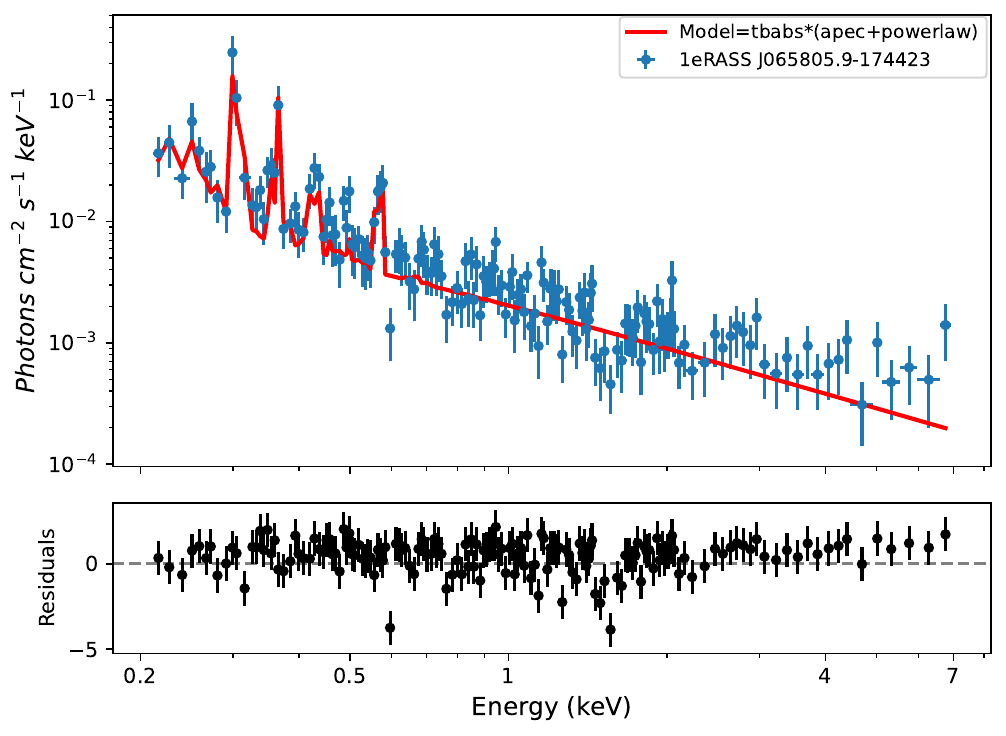}
      \includegraphics[width=0.4\textwidth]{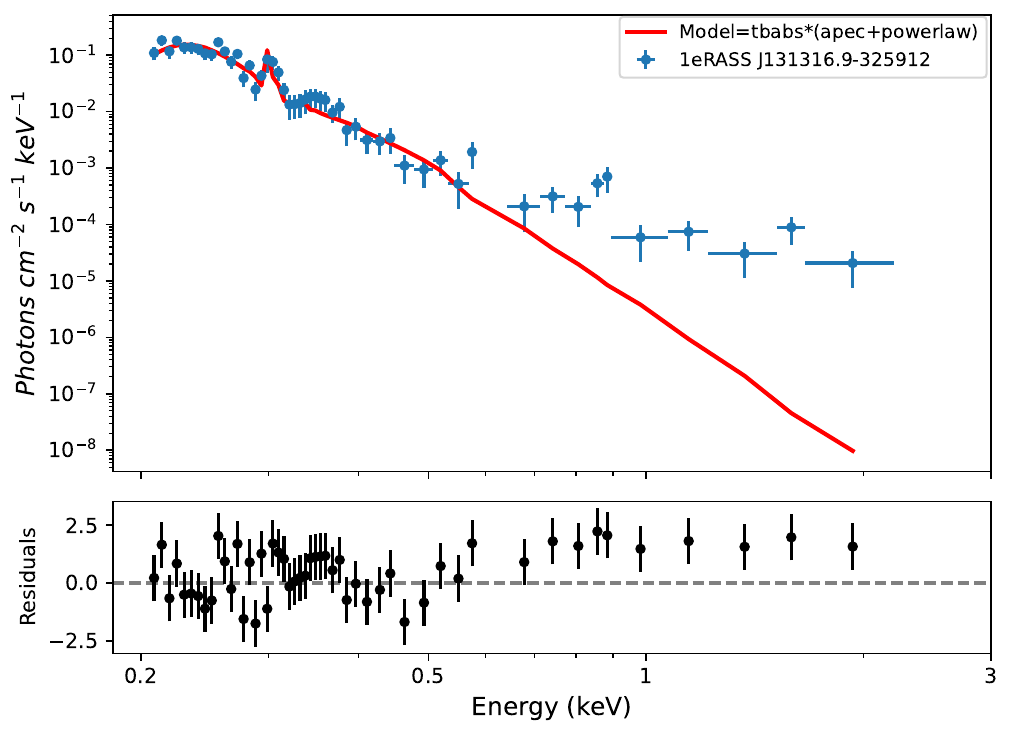}
      \includegraphics[width=0.40\textwidth]{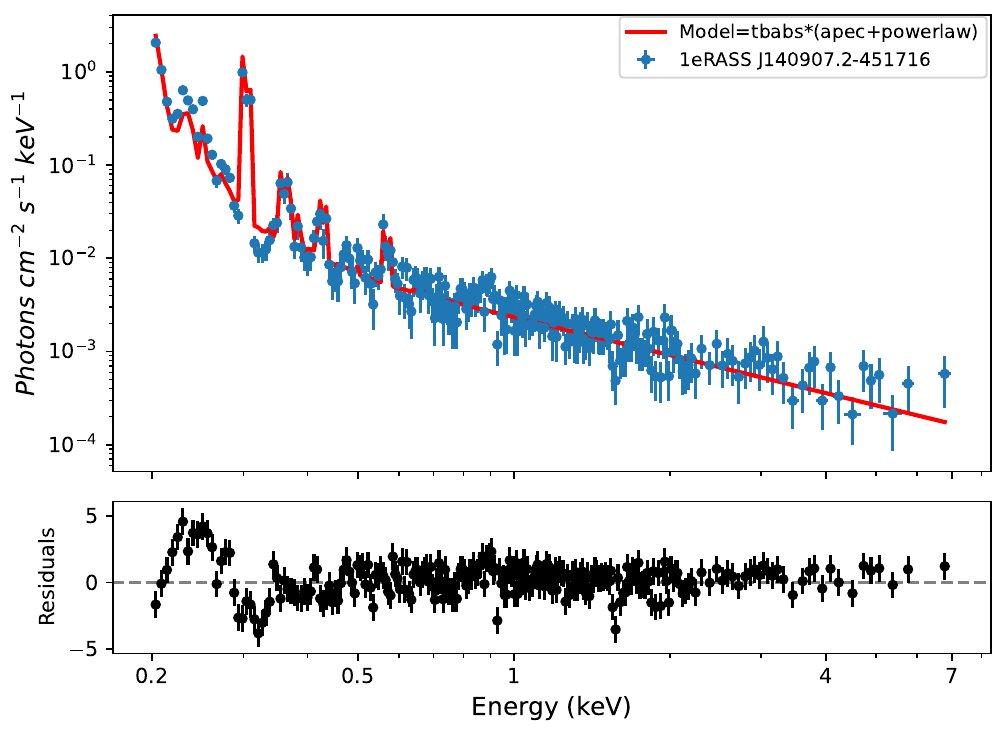}
      
    \caption{Individual eROSITA unfolded spectra of transients mentioned in the lower panel of Figure~\ref{fig:CV spectra}. Except "1eRASS J051541.4+010440" all other transients show good fit with either tbabs*(apec+powerlaw) or powerlaw models.}
    \label{fig:Lower Panel CVs}
\end{figure*}


\onecolumn

\newgeometry{left=1.3in, right=1.3in, top=1in, bottom=1in}
\setlength{\LTcapwidth}{\textwidth} 
\renewcommand{\arraystretch}{1.4}  
\setlength{\tabcolsep}{1.5pt}



\onecolumn

\begin{scriptsize}
\begin{longtable}{cccccccccccc}
\caption{This Table presents the properties of Galactic transients from the eRASS1 survey. The quantities in parentheses indicate measurement errors. Right Ascension (RA) and Declination (DEC) are based on Gaia DR2 coordinates; we've omitted error details for these coordinates as they are relatively small, on the order of milli-arcseconds. The spectral luminosity classification for the transients is sourced either from $\textit{Simbad}$ or derived from temperature and radius data in Gaia DR2. Period (P) information is obtained from Gaia DR3. $\frac{F_{\mathrm{2RXS}}}{F_{\mathrm{eRASS1}}}$ value for the sources, unique to eRASS1, represents an upper limit.} \\
\hline
\textbf{ID} & \textbf{RA} & \textbf{DEC} & \textbf{eRASS1 Flux} & \textbf{2RXS Flux} & \textbf{Distance} & \textbf{Luminosity} & \textbf{BP-RP} & \textbf{G}  & \textbf{Spec-Lum} & \textbf{P} & \textbf{\rule{0pt}{4ex}$\frac{F_{\rm 2RXS}}{F_{\rm eRASS1}}$} \\
\textbf{(1eRASS)} & \textbf{(Deg)} & \textbf{(Deg)} & \textbf{$10^{-14}$(erg s$^{-1}$ cm$^{-2}$)} & \textbf{$10^{-14}$(erg s$^{-1}$ cm$^{-2}$)} & \textbf{(pc)} & \textbf{$10^{29}$(erg s$^{-1}$)} & & & \textbf{Classification} & \textbf{(Hours)} & \\
\hline
\endfirsthead

\hline
\textbf{ID} & \textbf{RA} & \textbf{DEC} & \textbf{eRASS1 Flux} & \textbf{2RXS Flux} & \textbf{Distance} & \textbf{Luminosity} & \textbf{BP-RP} & \textbf{G}  & \textbf{Spec-Lum} & \textbf{P} & \textbf{\rule{0pt}{4ex}$\frac{F_{\rm 2RXS}}{F_{\rm eRASS1}}$} \\
\textbf{(1eRASS)} & \textbf{(Deg)} & \textbf{(Deg)} & \textbf{$10^{-14}$(erg s$^{-1}$ cm$^{-2}$)} & \textbf{$10^{-14}$(erg s$^{-1}$ cm$^{-2}$)} & \textbf{(pc)} & \textbf{$10^{29}$(erg s$^{-1}$)} & & & \textbf{Classification} & \textbf{(Hours)} & \\
\hline
\endhead

\hline
\endfoot

\endlastfoot
        \midrule
        \multicolumn{12}{c}{\textbf{CV}} \\ 
        \hline
          J002133.2-514233 & 5.38841 & -51.70967 & 105(9) & -- & 315(4) & 125(11) & 1.55 & 15.4 & K V & -- & $<$0.32 \\ 
         J003831.0-640312 & 9.62936 & -64.05381 & 53(6) & -- & 333(18) & 70(11) & 0.37 & 18.3 & -- & -- & $<$0.64 \\ 
         J004214.2-560920 & 10.55964 & -56.15557 & 91(8) & -- & 621(44) & 417(69) & 0.60 & 17.6 & -- & -- & $<$0.37 \\ 
         J014100.5-675326 & 25.25219 & -67.89088 & 265(11) & 2946(107) & 131(1) & 54(3) & 0.76 & 17.2 & A  V & -- & 11.12 \\ 
         J031130.8-315250 & 47.87888 & -31.88083 & 221(10) & -- & 274(9) & 198(16) & 0.10 & 16.3 & A V  & -- & $<$0.15 \\ 
         J033528.6-254421 & 53.86947 & -25.73939 & 104(7) & 493(174) & 240(4) & 71(5) & 0.59 & 16.7 & M  V & -- & 4.74 \\ 
         J051300.0-252323 & 78.25017 & -25.38975 & 150(9) & -- & 623(61) & 695(142) & 1.73 & 18.4 & -- & -- & $<$0.23 \\
\end{longtable}
\end{scriptsize}

\twocolumn






\bsp	
\label{lastpage}
\end{document}